\documentclass[useAMS,usenatbib]{mn2e}
\usepackage{times}
\input{epsf}
\usepackage{epsfig}
\usepackage{amssymb}

\def\sax1808{SAX~J1808.4--3658}
\def\1751{XTE~J1751--305}
\def\Rozanska{R{\'o}{\.z}a{\'n}ska}

\def\rinf{R_{\rm \infty}}
\def\rspot{R_{\rm spot}}

\def\Te{T_{\rm e}}

\def\rg{R_{\rm s}}
\def\Ledd{L_{\rm Edd}}
\def\taut{\tau_{\rm T}}
\def\be{\begin{equation}}
\def\ee{\end{equation}}
\def\beq{\begin{eqnarray}}
\def\eeq{\end{eqnarray}}
\def\msun{{\rm M_\odot}}
\def\rmd{{\rm d}}
\def\NH{N_{\rm H}}
\def\alphac{\alpha_{\rm c}}

\title[Physics of accretion in XTE~J1751--305]
{Physics of accretion in the millisecond pulsar
XTE~J1751--305}

\author[M. Gierli\'nski and J.Poutanen]
{Marek~Gierli\'nski$^{1, 2, 3}$\thanks{E-mail:
marek.gierlinski@durham.ac.uk  (MG), juri.poutanen@oulu.fi (JP)}
and Juri Poutanen$^{3}$\footnotemark[1]
\thanks{Corresponding Fellow, NORDITA, Copenhagen} \\
$^1$Department of Physics, University of Durham, South Road,
Durham DH1 3LE,
UK\\
$^2$Astronomical Observatory, Jagiellonian University, Orla 171,
30-244
Krak{\'o}w, Poland\\
$^3$ Astronomy Division, P.O.Box 3000, FIN-90014 University of
Oulu, Finland}

\date{Submitted to MNRAS}
\pagerange{\pageref{firstpage}--\pageref{lastpage}} \pubyear{2005}

\begin{document}

\topmargin = -0.5cm

\maketitle

\label{firstpage}

\begin{abstract}
We have undertaken an extensive study of X-ray data from the
accreting millisecond pulsar XTE J1751--305 observed by {\it
RXTE\/} and {\it XMM-Newton\/} during its 2002 outburst. In all
aspects this source is similar to a prototypical millisecond
pulsar SAX J1808.4--3658, except for the higher peak luminosity of
13 per cent of Eddington, and the optical depth of the hard X-ray
source larger by factor $\sim$2.
Its broad-band X-ray spectrum can be modelled by three components.
We interpret the two soft components as thermal emission from a
colder ($kT \sim 0.6$ keV) accretion disc and a hotter ($\sim 1$
keV) spot on the neutron star surface. We interpret the hard
component as thermal Comptonization in plasma of temperature
$\sim$40 keV and optical depth of $\sim$1.5 in a slab geometry.
The plasma is heated by the accretion shock as the material
collimated by the magnetic field impacts on to the surface. The
seed photons for Comptonization are provided by the hotspot, not
by the disc. The Compton reflection is weak and the disc is
probably truncated into an optically thin flow above the
magnetospheric radius.
Rotation of the emission region with the star creates an almost
sinusoidal pulse profile with rms amplitude of 3.3 per cent. The
energy-dependent soft phase lags can be modelled by two pulsating
components shifted in phase, which is naturally explained by a
different character of emission of the optically thick spot and
optically thin shock combined with the action of the Doppler boosting.
The observed variability amplitude constrains the hotspot to lie
within 3--4$\degr$ of the rotational pole.
We estimate the inner radius of the
optically thick accreting disc of about 40 km. In that case,
the absence of the emission from the antipodal spot,
which can be blocked by the accretion disc, gives the inclination
of the system to be   $\gtrsim 70\degr$.
\end{abstract}


\section{Introduction}
\label{sec:introduction}

Millisecond radio pulsars are believed to be spun-up to their fast
rotational speeds during the phase of accretion from their
companion stars (for a review see e.g. \citealt{b95}). Though
fast radio pulsars have been known for over 20 years,
accretion-powered millisecond pulsars had proven to be an elusive
missing link for a long time. \sax1808, the first millisecond
pulsar (spin period $P_s$ = 2.5 ms) in a low-mass X-ray binary
(LXMB) was discovered in 1998 \citep{wk98a} by
{\it Rossi X-ray Timing Explorer} ({\it RXTE}). Recent discoveries
of \1751 ($P_s$ = 2.3 ms; \citealt{mar02}, hereafter M02),
XTE J0902--314 ($P_s$ = 5.4 ms; \citealt{gal02}), XTE J1807--294 ($P_s$ =
5.3 ms; \citealt*{mar03}), XTE J1814--338 ($P_s$ =
3.2 ms; \citealt{ms03}),  and IGR J00291+5934 ($P_s$ =
1.67 ms; \citealt{eck04,mss04}) have brought the total number of
currently known accreting millisecond pulsars to six.

\citet*[][henceforth GDB02]{gdb02} analysed
energy spectra of \sax1808 from the 1998 outburst. The X-ray
spectral shape remained almost constant throughout the outburst
and could be fitted by a blackbody and thermal Comptonization.
Weak Compton reflection ($\Omega/2\pi \sim 0.1$) was also
detected. GDB02 interpreted the blackbody as emission from the
heated spot on the neutron star, and the Comptonization taking
place in an accretion column, as the material collimated by the
magnetic field impacts on to the neutron star surface. The
phase-resolved X-ray spectra showed that pulsation of these two
components is shifted in phase. \citet[][henceforth PG03]{pg03}
 applied the full relativistic treatment to the
energy-resolved pulse profiles of \sax1808 and studied the effects
of different angular  distribution of the emission pattern from the blackbody and
Comptonization components. They estimated the neutron star radius to be
 $\sim$ 6.5 and $\sim$ 11 km for a 1.2 and 1.6 M$_\odot$
neutron star, respectively.

\1751 is in a very tight binary, with the orbital period of only 42 minutes
and the mass function of the pulsar $1.3\times10^{-6}$ M$_\odot$, which gives
a minimum mass for the companion of about 0.014 M$_\odot$ (M02). The distance,
$D$, to the source is not known, though as it might be related to the Galactic
centre, therefore we assume $D \sim$ 8.5 kpc. M02 estimated the distance to be
greater than 7 kpc, using indirect arguments.

In this paper we examine the X-ray spectra of \1751 from its April 2002
outburst, observed by {\it XMM-Newton\/} and {\it RXTE}. In Section
\ref{sec:phase-averaged} we study the (phase-averaged) energy spectra, in
particular the broad-band 0.7--200 keV simultaneous spectrum from both
instruments. We model them by the blackbody and Comptonization from hotspot and
accretion shock, respectively.
We also show that another soft component, from the accretion disc,
is required. In Sections \ref{sec:pulse_profiles} and \ref{sec:phase-resolved}
we study phase-resolved spectra and interpret them in terms of two
independently pulsating spectral components. In Section \ref{sec:discussion}
we discuss our results and derive constraints on the accretion flow geometry
which occur to be very similar to those found for \sax1808 (GDB02).

\section{Observations}
\label{sec:observations}

\subsection{\it RXTE}

{\it RXTE\/} observed \1751 from 3$^{\rm rd}$ to 30$^{\rm th}$ of
April 2002. We reduced these data using {\sc ftools} version 5.3.
Proportional Counter Array (PCA) units 1 and 4 were switched off
for most of this period. As unit 0 lost its propane layer in May
2000 and its calibration is uncertain, we decided to use the data
from units 2 and 3 only. For each of the observations with the
unique identifier (obsid) we extracted energy spectra in 3--20 keV
band. Analogous Crab spectra from the same period show features at
a level of ~1 per cent, most likely of the instrumental origin.
Therefore, we applied 1 per cent systematic errors in each of the
PCA energy channels in addition to statistical errors.

In addition to the PCA data we also extracted 20--200 keV
High-Energy X-ray Timing Experiment (HEXTE) spectra from both
detector clusters. During observation of April 3 HEXTE did not
perform `rocking' and did not record background, so we could not
use these data. Table \ref{tab:obslog} contains log of the {\it
RXTE\/} observations analysed in this paper.

\1751 is only ~2$^\circ$ away from the Galactic centre, so its
X-ray spectrum is contaminated by the diffuse Galactic ridge
emission. Fortunately, {\it RXTE} performed several observations
in the direction of \1751 after the outburst (see fig.~1 in M02).
Between 20$^{\rm th}$ and 25$^{\rm th}$ of April PCA count rate
remained constant and minimal (there was a small outburst later
on), indicating its origin from the background. We accumulated
average PCA spectrum from this period (with average count rate of
$12.01\pm0.04$ s$^{-1}$ in detectors 2+3, all layers) and used it
as an additional background for outburst observations. Table
\ref{tab:obslog} contains PCA count rates corrected for this
effect. HEXTE spectra were not affected by the diffuse emission.

We extracted phase-resolved energy spectra from the PCA Event mode
files (configuration E\_125us\_64M\_0\_1s) with timing resolution
of 122 $\mu$s (except for the April 3 observation where we used
GoodXenon mode files, with resolution of 1 $\mu$s). We generated
folded light-curves in 16 phase bins, for each PCA channel, using
events from all layers of units 2 and 3. We have chosen beginning
of the phase ($\phi = 0$) at the bin with lowest 3--20 keV count
rate.  Photon arrival times were corrected for orbital movements
of the pulsar and the spacecraft, using ephemeris from M02.
Background files were created from standard models,
with addition of the April 20--25 spectrum to represent the
diffused emission. We also extracted power density spectra (PDS)
from the same event data files, but using units 0, 2 and 3. We
have created power density spectra in the $^1/_{128}$--512 Hz
frequency range from averaging fast Fourier transforms over 128-s
data intervals.

\subsection{\it XMM-Newton}

{\it XMM-Newton\/} observed \1751 on 2002 April 7, with the total
exposure of 33 ks. We reduced the data from European Photo Imaging
Camera PN (EPIC-pn) detector using {\sc sas} version 5.4.1. and
following guidelines in the MPE cookbook
(http://wave.xray.mpe.mpg.de/xmm/cookbook). The detector operated
in timing mode where spatial information was compressed into one
dimension. We extracted EPIC-pn spectrum from a stripe in raw CCD
coordinates RAWX (36,56), collecting single and double events
only. The background was extracted from the adjacent stripe in
RAWX (16,36).

All spectral analysis (both phase-resolved and phase-averaged) was
done using the {\sc xspec} 11.2 spectral package \citep{arn96}.
The error of each model parameter is given for a 90 per cent
confidence interval, except for the pulse profiles in
Sec.~\ref{sec:phase-resolved}, where we used 1$\sigma$ errors. The
relative normalization of the PCA and HEXTE instruments is
uncertain, so we allowed this to be an addition free parameter in
all spectral fits.

\begin{table*}
\centering \caption{Log of {\it RXTE} observations analysed in
this paper. Start and end times are in the UT days of April 2002.
Exposures are in seconds and (background subtracted) count rates
are in counts per second from detectors 2 and 3 of PCA and
separately from cluster 0 and 1 of HEXTE. During the first
observation HEXTE did not observe background.} \vspace{12pt}
\begin{tabular}{clccrrrrrr}
\hline
 & & & &  \multicolumn{2}{c}{PCA (2+3)} & \multicolumn {2}{c}{HEXTE 0} &  \multicolumn {2}{c}{HEXTE 1}  \\
No. & Obsid & Start & End & Exposure & Count rate & Exposure & Count rate & Exposure & Count rate\\
\hline

 1 & 70134-03-01-00  &  3.653 &   3.720 &  3168  & 250.5$\pm$0.6 & \multicolumn{2}{c}{N/A}  & \multicolumn{2}{c}{N/A} \\
 2 & 70131-01-01-00  &  4.644 &   4.912 & 10960  & 237.3$\pm$0.5 &  3520  &  18.7$\pm$0.2 &  3461  &  15.2$\pm$0.2 \\
 3 & 70134-03-02-00  &  5.162 &   5.179 &   720  & 226.5$\pm$0.8 &   257  &  19.8$\pm$0.9 &   263  &  14.4$\pm$0.8 \\
 4 & 70131-01-02-00  &  5.535 &   5.773 &  6384  & 214.8$\pm$0.5 &  2039  &  17.3$\pm$0.3 &  2012  &  14.0$\pm$0.2 \\
 5 & 70131-01-03-01  &  6.525 &   6.542 &  1456  & 190.0$\pm$0.6 &   425  &  16.4$\pm$0.6 &   431  &  12.3$\pm$0.5 \\
 6 & 70131-01-03-00  &  6.591 &   6.763 &  9200  & 185.9$\pm$0.4 &  2859  &  15.3$\pm$0.2 &  2791  &  12.0$\pm$0.2 \\
 7 & 70131-01-04-00  &  7.515 &   7.753 & 12416  & 163.3$\pm$0.4 &  3839  &  13.7$\pm$0.2 &  3809  &  10.7$\pm$0.2 \\
 8 & 70131-01-05-03  &  8.061 &   8.083 &  1168  & 162.4$\pm$0.6 &   432  &  14.3$\pm$0.7 &   410  &   8.9$\pm$0.6 \\
 9 & 70131-01-05-02  &  8.131 &   8.149 &   864  & 154.8$\pm$0.7 &   313  &  14.4$\pm$0.9 &   302  &  10.3$\pm$0.8 \\
10 & 70131-01-05-01  &  8.199 &   8.215 &   576  & 155.4$\pm$0.8 &   204  &  12.6$\pm$1.1 &   198  &   9.3$\pm$1.0 \\
11 & 70131-01-05-04  &  8.267 &   8.467 &  2368  & 147.7$\pm$0.5 &   754  &  13.8$\pm$0.5 &   749  &   9.4$\pm$0.4 \\
12 & 70131-01-05-000 &  8.505 &   8.743 & 12544  & 144.9$\pm$0.4 &  3897  &  13.4$\pm$0.2 &  3799  &   9.6$\pm$0.2 \\
13 & 70131-01-05-00  &  8.772 &   9.007 &  4752  & 143.1$\pm$0.4 &  1645  &  12.9$\pm$0.3 &  1586  &   8.9$\pm$0.2 \\
14 & 70131-01-06-00  &  9.435 &   9.601 &  8800  & 129.0$\pm$0.3 &  2831  &  12.6$\pm$0.2 &  2731  &   8.6$\pm$0.2 \\
15 & 70131-01-06-01  &  9.627 &   9.799 &  8464  & 126.1$\pm$0.3 &  2714  &  12.1$\pm$0.2 &  2660  &   8.0$\pm$0.2 \\
16 & 70131-01-07-00  & 10.418 &  10.525 &  6592  & 112.1$\pm$0.3 &  2093  &  10.4$\pm$0.3 &  2092  &   7.1$\pm$0.2 \\
17 & 70131-01-07-01  & 10.551 &  10.789 & 11552  & 108.6$\pm$0.3 &  3619  &   9.9$\pm$0.2 &  3573  &   6.8$\pm$0.2 \\
18 & 70131-01-08-000 & 11.409 &  11.713 & 15552  &  94.3$\pm$0.3 &  4840  &   8.9$\pm$0.2 &  4794  &   5.9$\pm$0.1 \\
19 & 70131-01-08-00  & 11.742 &  11.912 &  6704  &  91.2$\pm$0.3 &  2376  &   8.7$\pm$0.2 &  2334  &   5.5$\pm$0.2 \\
20 & 70131-01-09-000 & 12.333 &  12.638 &  9632  &  56.7$\pm$0.2 &  2934  &   5.9$\pm$0.2 &  2870  &   3.3$\pm$0.2 \\
21 & 70131-01-10-00G & 13.322 &  13.496 &  5120  &  13.7$\pm$0.2 &  1683  &   1.3$\pm$0.3 &  1686  &   0.6$\pm$0.2 \\

\hline
\end{tabular}
\label{tab:obslog}
\end{table*}

\section{Outburst}
\label{sec:outburst}

\begin{figure}
\begin{center}
\leavevmode \epsfxsize=7cm \epsfbox{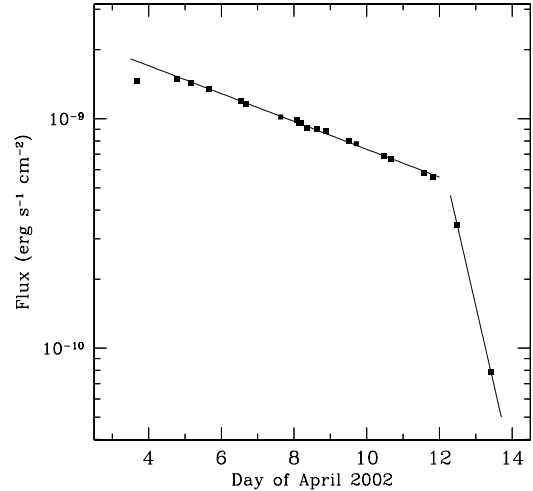}
\end{center}
\caption{3--20 keV unabsorbed flux of observations 1--21 (Table
\ref{tab:obslog}). The solid lines correspond to $F \propto
e^{-t/7.2^{\rm d}}$ and $F \propto e^{-t/0.63^{\rm d}}$.}
\label{fig:lightcurve}
\end{figure}

Fig.~\ref{fig:lightcurve} shows the evolution of the 3--20 keV unabsorbed flux
during the outburst (see also M02). The peak flux during observation 2 was
$1.5\times10^{-9}$ erg s$^{-1}$ cm$^{-2}$. Using a Comptonization model (see
Section~\ref{sec:phase-averaged}), we estimated the peak bolometric
X-ray/$\gamma$-ray flux of $3.1\times10^{-9}$ erg s$^{-1}$ cm$^{-2}$. For a
distance of 8.5 kpc (assuming the source to be close to the Galactic centre)
this corresponds to the bolometric luminosity of $2.7\times10^{37}$ erg
s$^{-1}$, or 13 per cent of Eddington luminosity $\Ledd$ for a 1.4 M$_\odot$ neutron
star. This is much brighter than \sax1808 with the estimated peak bolometric
luminosity of 2.2 per cent of  $\Ledd$ (GDB02).

The light curve of \1751 shows a striking similarity to that of
\sax1808 (see fig.~1 in \citealt*{g98}; see also M02).
 After the peak the flux declined exponentially, with
$e$-folding factor of 7.2 days (10 days in \sax1808), until it
reached a break after which the flux dropped suddenly with the
$e$-folding factor of $\sim$0.6 days (1.3 days in \sax1808). The
break in the light curve is most likely associated with the onset
of the cooling wave in the accretion disc (GDB02).

\section{Phase-averaged energy spectra}
\label{sec:phase-averaged}

\begin{table*}
\centering \caption{Summary of spectral models used in this paper.} \vspace{12pt}
\begin{tabular}{clp{9cm}}
\hline
Model & {\sc xspec} components & Description \\
\hline

DTH & {\sc wabs*(diskbb+thcomp)} & Multicolour disc and thermal
Comptonization of the disc photons.\\\\

DTF & {\sc wabs*(diskbb+thcomp)} & The same as DTH, but the seed
photon temperature is free and independent of the disc.\\\\

DBTH & {\sc wabs*(diskbb+bbodyrad+thcomp)} & Multicolour disc,
single-temperature blackbody and thermal Comptonization of the
blackbody photons. \\\\

DBTF & {\sc wabs*(diskbb+bbodyrad+thcomp)} & The same as DBTH, but
the seed photon temperature is free and independent of the blackbody.\\\\

DBPS & {\sc wabs*(diskbb+bbodyrad+compps)} & The same as DBTH,
with {\sc thcomp} replaced by {\sc compps}.\\

DBPF & {\sc wabs*(diskbb+bbodyrad+compps)} & The same as DBTF,
with {\sc thcomp} replaced by {\sc compps}.\\
\hline
\end{tabular}
\label{tab:models}
\end{table*}

\subsection{Physical picture and spectral models}

If the neutron star  magnetic field is strong enough,
the accreting material will follow magnetic field lines
and form a shock close to the  neutron star surface
\citep[for the  accretion geometry see fig. 12 of GDB02 and][]{rom04}.
The shock is  pinned down to
the stellar surface at luminosity of a few per cent of $\Ledd$ \citep{bs76,ls82}.
Gravitational energy is  dissipated in the shock and is
transferred from the protons to the electrons by Coulomb collisions.
The main cooling mechanism
is Comptonization of soft photons provided by the neutron star.
The stellar surface is heated under the shock, while
the  hard  X-rays,  produced  in the  shock,  can  also irradiate
the surrounding surface, so that the black body emission region
can cover a somewhat larger area.

Such a physical picture is in agreement with
observations of \sax1808 (GDB02, PG03).
We presume that \1751 is a similar source to \sax1808.
Thus we expect emission
from the hotspot on the neutron star surface, which we model as a blackbody.
We also expect emission from the accretion shock, which we model by thermal
Comptonization. We consider two possible
geometries of the shock and hotspot, shown in Fig.~\ref{fig:shockgeom}. For
the phase-averaged fits we assume that the angle between the rotation axis and
the magnetic pole is small, so that the slab model of the shocked region
viewed at a fixed angle is justified.
We will show later in Sec.~\ref{sec:constampl} that this assumption
is feasible. In addition to the above components we might expect thermal
emission from the accretion disc and Compton reflection from the disc and/or
neutron star surface.

For the spectral description we use the models summarized in
Table \ref{tab:models}. Their spectral components are as
follows. {\sc bbodyrad} is a single-temperature blackbody,
normalized to its apparent area (at a given distance), which
represents the emission from a hotspot at a neutron star
surface. {\sc diskbb} is a multicolour (accretion) disc
model \citep{mit84}, normalized to the inner disc radius (at
a given distance). We correct the inner disc radius for the
spectral hardening with factor $f_{\rm col}$ = 1.8
\citep{st95} and for the torque-free inner boundary
condition, with factor $\zeta$ = 0.37 \citep{g99}: $R_{\rm
in} = f^2_{\rm col} \zeta R_{\rm in, apparent}$. The exact
value of the hardening factor is not well known, but recent
numerical accretion disc simulations place it between 1.8
and 2.0, increasing with luminosity (Davis et al. 2004).
Additional uncertainty comes from general and special
relativistic corrections \citep[see e.g.][]{c75,zcc97}.
Later in this paper we discuss the effects of the disc with
the continuous torque through the inner radius.

{\sc thcomp} is a thermal Comptonization model, using an approximate solution
of the \citet{kom56} equation \citep{zjm96}. It is parameterized by the
asymptotic power-law photon index, $\Gamma$, electron temperature, $\Te$ and
seed photon temperature, $T_{\rm seed}$. {\sc compps} is another
Comptonization model \citep{ps96}. It finds an exact numerical solution of the
Comptonization problem explicitly considering successive scattering orders. It
also allows for various hot plasma geometries. For a slab geometry of
the emission region, this model is parameterized by
the optical depth of the scattering medium, $\tau$, electron temperature,
$\Te$,  seed photon temperature, $T_{\rm seed}$, and the inclination angle
to the slab normal. The interstellar
absorption was described by model {\sc wabs} with the hydrogen column density
$\NH$ as  a parameter.

\subsection{\it XMM-Newton}
\label{sec:pn_fits}

\citet{m03} successfully fitted EPIC-pn spectrum of \1751 with a simple
phenomenological model of a blackbody and power law. We confirmed that this
model gave a very good reduced $\chi^2/\nu$ = 2013/1859, indeed. There are,
however, two fundamental issues here. Firstly, the most probable radiative
process that gives rise to emission above $\sim$1 keV is thermal
Comptonization which can be described by a power law  in a narrow energy band only.
At energies around seed photon temperature, the Comptonization spectrum has a
low-energy cutoff. Secondly, the best-fitting spectral index of the power law
($\Gamma = 1.44\pm0.01$) is in strong disagreement with {\it RXTE\/} data,
where the spectrum is much softer, $\Gamma \sim$ 1.8--1.9 (see below).

\begin{figure}
\centerline{\epsfig{file=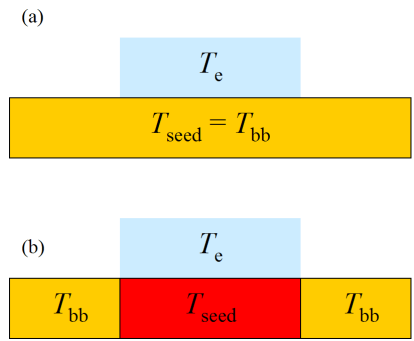,width=5.5cm} }
\caption{Schematic geometries of the emission region, consisting of the
accretion shock above, and the hotspot on the surface of the neutron star. (a)
Geometry corresponds to a small temperature gradient in the hotspot, i.e.
the case where seed photons have the same temperature
as the photons in the observed black body component (models DBTH, DBPS). (b)
Geometry corresponds to a large temperature gradient where
seed photons are hotter than the illuminated region of the
neutron star surface around the shock, which gives rise to the observed black body
emission (models DBTF, DBPF). The emitting region corresponds to a point where
the accretion column impacts on the surface (see fig.~12 in GDB02).
} \label{fig:shockgeom}
\end{figure}

Therefore, we tried a physically motivated model. Instead of the
power law we used {\sc thcomp}, assuming that the seed photons
were from the blackbody component. This model gave rather poor fit
with $\chi^2/\nu$ = 2754/1859 and broad residuals implying wrong
shape of the continuum. In the second attempt we allowed the seed
photons for Comptonization to be independent of the blackbody
component. This fit was much better ($\chi^2/\nu$ = 2044/1858),
but the large apparent area of the blackbody, $A_{\rm bb} \sim$
1400 ($D$/8.5 kpc)$^2$ km$^2$, ruled out its origin from the
neutron star surface. Instead, it rather originated from the
accretion disc, so we replaced the blackbody by the multicolour
disc (model DTF in Table \ref{tab:models}). The results are shown
in Table \ref{tab:pn_fits}. The seed photons for Comptonization
($kT_{\rm seed} = 0.68_{-0.01}^{+0.02}$ keV) were much hotter than
the disc photons ($kT_{\rm disc} = 0.38_{-0.02}^{+0.03}$ keV) and
they might have originated from the neutron star surface (see
\citealt*{gd02}; GDB02). We tried to model the surface
emission by adding a single-temperature blackbody component with
its temperature tied to the seed photons (see
Fig.~\ref{fig:shockgeom}a). Model DBTH improved the fit by
$\Delta\chi^2 = 28$ over model DTF, with one degree of freedom
less (Table \ref{tab:pn_fits}). The apparent area of this
blackbody, $A_{\rm bb} = 36_{-4}^{+3}$ ($D$/8.5 kpc)$^2$ km$^2$,
was consistent with a hotspot on the neutron star surface.

\begin{table}
\centering \caption{Best-fitting parameters of the models applied
to EPIC-pn data. Models are described in Table \ref{tab:models}.
Inner disc radius $R_{\rm in}$ and apparent area of the blackbody
$A_{\rm bb}$ were calculated assuming distance of 8.5 kpc. {\sc
thcomp} temperature was fixed at 30 keV.} \vspace{12pt}
\begin{tabular}{lrr}
\hline
Model & DTF & DBTH \\
\hline
$\NH$ (10$^{22}$ cm$^{-2}$) & 1.04$_{-0.03}^{+0.02}$ & 1.01$_{-0.02}^{+0.01}$\\
$kT_{\rm disc}$ (keV) & 0.38$_{-0.02}^{+0.03}$ & 0.51$_{-0.03}^{+0.09}$\\
$R_{\rm in} \sqrt{\cos i}$ (km) & 16.5$_{-2.6}^{+2.5}$ & 10.0$_{-2.2}^{+1.5}$\\
$kT_{\rm bb}$ (keV) & - & 0.89$_{-0.05}^{+0.09}$\\
$A_{\rm bb}$ (km$^2$) & - & 36$_{-4}^{+3}$\\
$\Gamma$ & 1.90$_{-0.01}^{+0.02}$ & 1.72$_{-0.14}^{+0.07}$\\
$kT_{\rm seed}$ (keV) & 0.68$_{-0.01}^{+0.02}$ & $= kT_{\rm bb}$\\
$\chi^2/\nu$ &  2039.5/1858 & 2011.9/1857 \\
\hline
\end{tabular}
\label{tab:pn_fits}
\end{table}

There were no distinct features around 7 keV in the residuals, and
we did not detect Compton reflection at a statistically
significant level in the EPIC-pn spectrum. The best-fitting
amplitude of reflection was only $\Omega/2\pi =
0.01_{-0.01}^{+0.06}$ giving fit improvement of $\Delta\chi^2$ =
1.9 with two degrees of freedom less.

\subsection{Simultaneous {\it XMM-Newton} and {\it RXTE}}
\label{sec:pn-xte_fits}

\begin{figure}
\begin{center}
\leavevmode \epsfxsize=7cm \epsfbox{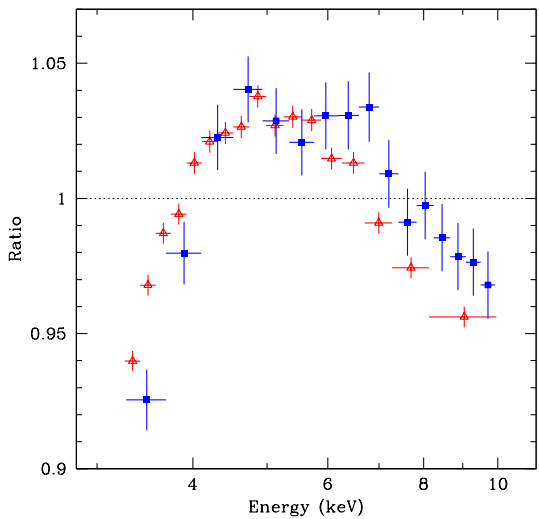}
\end{center}
\caption{EPIC-pn/PCA cross-calibration. The figure shows the
residuals (data-to-model ratio) of the EPIC-pn (triangles) and PCA
(squares) spectra to a power-law model. The EPIC-pn spectrum was
rebinned for presentation only.} \label{fig:cal}
\end{figure}

\begin{figure*}
\begin{center}
\leavevmode \epsfxsize=12cm \epsfbox{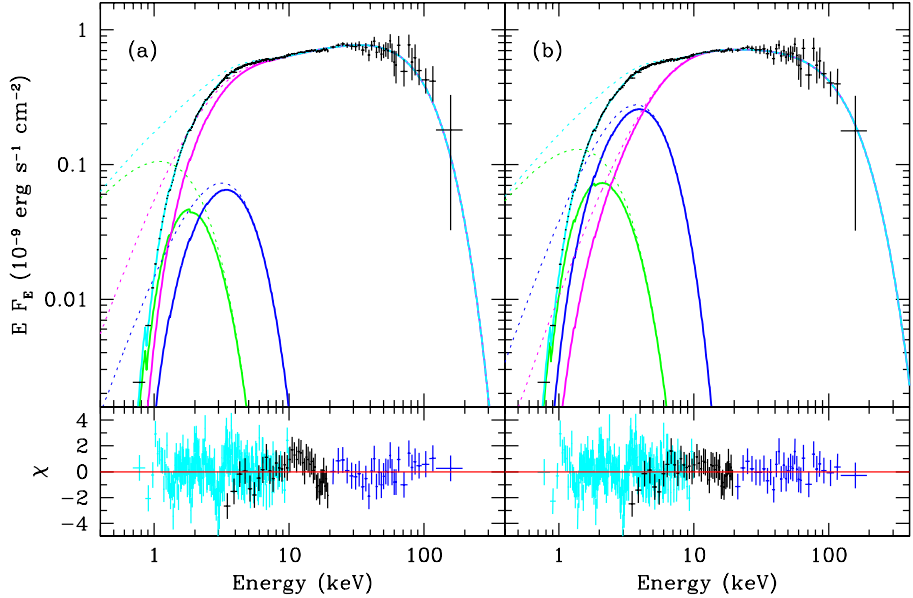}
\end{center}
\caption{Broad-band spectrum of \1751 from EPIC-pn, PCA and HEXTE.
The models consists of the multicolour disc (green curve),
single-temperature blackbody (blue curve), and thermal
Comptonization of the blackbody photons (magenta curve). The
dotted curves show unabsorbed spectral components. Panel (a)
shows model DBTH, corresponding to the shock geometry shown in
Fig.~\ref{fig:shockgeom}a. Panel (b) shows model DBTF, with
geometry as in Fig.~\ref{fig:shockgeom}b. See Table
\ref{tab:models} for model description and Table \ref{tab:px_fits}
for best fitting parameters. The lower panels shows the residuals
of EPIC-pn (cyan), PCA (black) and HEXTE (blue).}
\label{fig:spec_px}
\end{figure*}

Encouraged by the results from {\it XMM-Newton} we extended the observed
bandwidth by adding an {\it RXTE\/} spectrum. Observation 6 from Table
\ref{tab:obslog} was simultaneous with EPIC-pn data. It overlapped with a
small fraction of the EPIC-pn exposure only, but there was very little
spectral variability throughout this observation, so spectra from both
instruments should be the same. We also checked the cross-calibration between
the two instruments to look for possible response discrepancies. We fitted
both spectra in the overlapping 3--10 keV band by a power law. The spectral
indices were in good agreement, $\Gamma = 1.660\pm0.005$ and $1.667\pm0.015$
for EPIC-pn and PCA, respectively. The residuals were also quite similar (see
Fig.~\ref{fig:cal}), indicating nearly identical spectral shape. We decided
that the broad-band analysis of the simultaneous {\it XMM-Newton} and {\it
RXTE\/} was feasible. We fitted the EPIC-pn spectrum and PCA/HEXTE observation
6 together with a sequence of models, which are summarized in Table
\ref{tab:models}. We allowed for the relative normalization of these
instruments to be free and used PCA normalization for flux calculations. The
relative normalizations of EPIC-pn, HEXTE clusters 0 and 1 with respect to PCA
were 0.64, 0.57 and 0.74, respectively, for our best fit. The fit results are
shown in Table \ref{tab:px_fits} and two of the fitted spectra in
Fig.~\ref{fig:spec_px}.

\begin{table*}
\centering \caption{Best-fitting parameters of several models
applied to EPIC-pn/PCA/HEXTE simultaneous data. The models are
summarized in Table \ref{tab:models}. Inner disc radius $R_{\rm
in}$ and apparent area of the blackbody $A_{\rm bb}$ are
calculated assuming distance of 8.5 kpc. The spectrum and
components of the model DBTH are shown in Fig.~\ref{fig:spec_px}. }
\vspace{12pt}
\begin{tabular}{lrrrrrr}
\hline
Model & DTH &  DTF & DBTH & DBTF & DBPS & DBPF\\
\hline
$\NH$ (10$^{22}$ cm$^{-2}$) & 0.926$_{-0.004}^{+0.005}$ & 1.050$_{-0.019}^{+0.033}$ & 1.015$_{-0.015}^{+0.020}$ & 1.000$_{-0.011}^{+0.023}$ & 1.012$_{-0.017}^{+0.018}$ & 0.998$_{-0.013}^{+0.014}$\\
$kT_{\rm disc}$ (keV) & 1.13$\pm$0.04 & 0.35$\pm0.02$ & 0.46$\pm0.03$ & 0.58$\pm$0.04 & 0.48$\pm$0.04 & 0.61$\pm$0.05\\
$R_{\rm in} \sqrt{\cos i}$ (km) & 3.6$\pm$0.2 & 25.2$_{-2.7}^{+5.7}$ & 14.9$_{-1.9}^{+2.7}$ & 10.4$_{-1.1}^{+1.8}$ & 14.0$\pm$2.0 & 9.8$_{-1.2}^{+1.7}$\\
$kT_{\rm bb}$ (keV) & - & - & 0.81$\pm0.03$ & 0.95$\pm$0.04 & 0.84$\pm$0.03 & 1.00$\pm$0.05\\
$A_{\rm bb}$ (km$^2$) & - & - & $49\pm4$ & 95$_{-9}^{+14}$ & 39$_{-5}^{+11}$ & 91$_{-12}^{+16}$\\
$\Gamma$ & 1.78$\pm$0.01 & 1.86$\pm0.01$ & 1.81$\pm0.01$ & 1.94$\pm$0.05 & - & - \\
$\tau$ & - & - & - & - & 1.93$_{-0.15}^{+0.23}$ & 1.47$_{-0.14}^{+0.26}$\\
$k\Te$ (keV) & 22$_{-2}^{+3}$ & 32$_{-5}^{+8}$ & $25_{-3}^{+4}$ & 42$_{-10}^{+70}$ & 29$\pm$4 & 36$_{-5}^{+3}$\\
$kT_{\rm seed}$ (keV) & = $kT_{\rm disc}$ & 0.64$\pm$0.01 & = $kT_{\rm bb}$ & 1.75$\pm$0.18 & =$kT_{\rm bb}$ & 2.2$_{-0.3}^{+0.1}$\\
$A_{\rm seed}$ (km$^2$) & - & - & - & - & 790$_{-140}^{+50}$ & 20$_{-7}^{+17}$\\
$\chi^2/\nu$ & 2271.8/1984 & 2208.1/1983 &  2154.3/1982 & 2127.0/1981 & 2141.2/1982 & 2125.5/1981\\
\hline
\end{tabular}
\label{tab:px_fits}
\end{table*}

First, we tested the possibility that the seed photons for
Comptonization came from the accretion disc (model DTH). The model
fitted the data fairly well (reduced $\chi^2/\nu = 1.15$). It
required however a very small inner radius of the disc, $R_{\rm
in} \sqrt{\cos i} = 3.6\pm0.2$ ($D$/8.5 kpc) km
(here $i$ is the disc inclination angle), smaller than the
expected neutron star radius.

The model DTF, where the seed photons for Comptonization were
independent of the disc temperature, gave a good fit to the
EPIC-pn data in Section~\ref{sec:pn_fits}. We applied the same
model to the broad-band spectrum (with free $\Te$). It improved
the fit by $\Delta\chi^2 = 63.7$ (at 1983 d.o.f.) with respect to
DTH. The disc temperature was much lower and its inner disc radius
much larger, making it consistent with an overall picture of a
millisecond accreting pulsar (GDB02) or, more generally, an atoll
source in the island state \citep[e.g.][]{b01}.

Adding the blackbody source of the seed photons (presumably
neutron star surface) as an explicit component (model DBTH)
improved the fit further by  $\Delta\chi^2 = 53.8$ (at 1982
d.o.f.). The apparent area of the blackbody component, $A_{\rm bb}
= 49\pm4$ ($D$/8.5 kpc)$^2$ km$^2$, was consistent with a hotspot
on the neutron star surface. However, this component did not
produce enough photons for Comptonization. The Comptonization
spectrum of $\Gamma \sim 1.8$ and $k\Te \sim 25$ keV requires a
similar luminosity in the seed and Comptonized photons, while the
observed blackbody component was $\sim$ 30 times weaker than {\sc
thcomp}. This means that majority of the seed photons were not
visible, e.g. because they were covered by the Comptonizing
accretion column (Fig.~\ref{fig:shockgeom}b). In fact, spectral
fits with {\sc compps} (see below) showed that the actual area of
the seed photons source greatly exceeded the expected area of the
neutron star.

Therefore, we considered one more model (DBTF), which included the
disc emission, the blackbody and Comptonization of unseen seed
photons of much higher temperature (see
Fig.~\ref{fig:shockgeom}b). This model produced a very good fit,
improving previous results by $\Delta\chi^2$ = 27.3. It gave high
seed photon temperature of $kT_{\rm seed} = 1.75\pm0.18$ keV and
much softer spectral index of $\Gamma = 1.94\pm0.05$. As we show
below in this section, the seed photons in this model were
consistent with the hotspot on the neutron star.

Next, we tested how sensitive the above results were to the choice
of a particular model. We replaced {\sc thcomp} by another
Comptonization code, {\sc compps}. It computes the exact numerical
solution of the radiative transfer equation for Comptonization in
a given geometry and allows finding the normalization (or apparent
area) of the blackbody seed photons. We chose a geometry in which
a slab of hot plasma is irradiated from below by the blackbody
seed photons. The parameter $\tau$ in the model is the vertical
optical depth of the slab, so the actual line-of-sight optical
depth depends on the inclination angle $i$. Unlike {\sc thcomp},
the unscattered seed photons transmitted through the slab are
already included in {\sc compps} model, so any additional
blackbody component in the model corresponds to photons not
entering the hot plasma. This geometry is consistent with an
accretion shock above the hotspot on the neutron star surface
(GDB02; PG03).

For this model we assumed that the inclination angle of the slab with respect
to the observer was equal to the inclination of the system, $i = 60^\circ$.
This is, certainly, an approximation, as the accretion shock rotates with the
neutron star surface. However, as we show later in this paper (see Section
\ref{sec:geometry} and Fig.~\ref{fig:incldelta}) the magnetic inclination
$\theta$ is small, $\la$ 10$^\circ$, so the angle at which we see the shock
doesn't vary much.

In the model DBPS, corresponding to the geometry in
Fig.~\ref{fig:shockgeom}(a), the temperature of the seed photons
was tied to the blackbody component. The apparent area of the
source of the seed photons was very large, $A_{\rm seed} =
790_{-140}^{+50}$ ($D$/8.5 kpc)$^2$ km$^2$. This definitely could
not be a hotspot on the surface, unless the distance to the source
is less than $\sim$3 kpc (but see Section \ref{sec:distance}).

A fit with seed photons independent of the blackbody temperature
(model DBPF corresponding to a geometry in
Fig.~\ref{fig:shockgeom}b) gave better $\chi^2$, but more
importantly, yielded much smaller apparent area of the seed
photons of $20_{-7}^{+17}$ ($D$/8.5 kpc)$^2$ km$^2$. On the other
hand, the apparent area of the (independent) blackbody component
was significantly larger, $A_{\rm bb} \approx 90$ ($D$/8.5
kpc)$^2$ km$^2$.

The great advantage of {\sc compps} model is that it computes
angle-dependent spectrum. For an intermediate optical depth of the
slab of $\tau \sim$ 1.5--2.0 the overall spectral shape of the
Comptonized continuum depends only weakly on the observer's angle
with respect to the slab. The main effect is the amount of
unscattered seed photons seen at different angles, as it is very
sensitive to the actual optical depth along the line of sight,
$\tau/\cos i$. So far, we have assumed quite large inclination
angle of 60$^\circ$. To check the angular dependence we repeated
the fits with models DBPS and DBPF at 30$^\circ$. The DBPS fit was
dramatically worse by $\Delta\chi^2$ = 75.4. Smaller angle gave
larger, not consistent with the observed spectrum, amount of
unscattered seed photons. The DBPF fit, with more freedom of
shaping the low-energy part of the spectrum gave the fits worse by
$\Delta\chi^2$ = 7.1 and optical depth of $\tau = 2.54\pm0.15$,
larger than at 60$^\circ$ by expected factor of $\cos 30^\circ /
\cos 60^\circ$. The actual line-of-sight optical depth to the seed
photons was about 2.9, so the observed unscattered fraction of the
seed photons was about 5 per cent. This explains why {\sc thcomp},
with no unscattered blackbody photons included in the model, could
fit the data well. When inclination was allowed to be free, the
best-fitting values were $i=79_{-35}^{+11}$ and $65_{-19}^{+10}$
deg, for DBPS and DBPF, respectively.

No reflection was significantly detected in any of these models.
We found only upper limits on $\Omega/2\pi$ of 0.06 and 0.41, from
DBTF and DBPF, respectively. The first model incorporated
reflection with a self-consistent iron line \citep*{zds98}, while
the second one included an independent {\sc diskline} model, with
one more free parameter and a less strict constrain.

\begin{figure}
\begin{center}
\leavevmode \epsfxsize=8.5cm \epsfbox{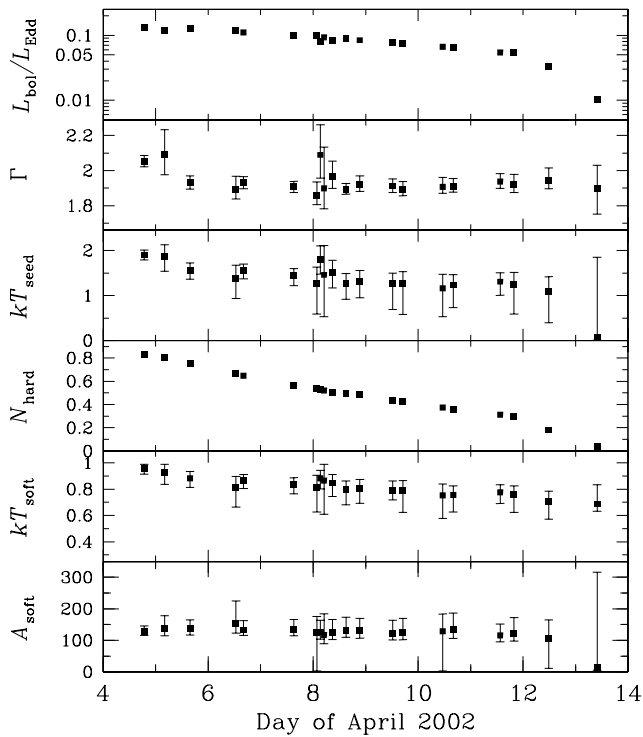}
\end{center}
\caption{Fitting results of the blackbody plus thermal
Comptonization model (DBTF from Table \ref{tab:models}, without
disc component) to the outburst {\it RXTE\/} spectra. The electron
temperature was fixed at 42 keV, the absorption column at
$1.0\times10^{22}$ cm$^{-2}$. $\Gamma$ is the asymptotic photon
spectral index, $kT_{\rm seed}$ (keV) is the seed photon
temperature, $N_{\rm hard}$ ($10^{-9}$ erg  s$^{-1}$ cm$^{-2}$) is
the normalization of the Comptonized component at 10 keV, $kT_{\rm
soft}$ (keV) and $A_{\rm soft}$ (km$^2$) are the temperature and
apparent area of the soft (blackbody) component, respectively. The
fraction of Eddington luminosity, $L_{\rm bol}/\Ledd$, and
$A_{\rm soft}$ are calculated for a distance of 8.5 kpc. }
\label{fig:pars}
\end{figure}

The total unabsorbed bolometric flux from model DBTF was $F_{\rm
bol} = 2.9\times10^{-9}$ erg s$^{-1}$ cm$^{-2}$, which corresponds
to luminosity $L_{\rm bol} = 2.5\times10^{37}$ erg s$^{-1}$ or
$0.12 \Ledd$ for a 1.4~M$_\odot$ neutron star at a distance
of 8.5 kpc. The contribution to the total luminosity from the disc
and blackbody components was 9 and 13 per cent, respectively.

\subsection{{\it RXTE\/} outburst}
\label{sec:xte_fits}

Finally, we analysed the {\it RXTE\/} spectra from Table
\ref{tab:obslog} covering the whole outburst (observations 2--21).
For spectral fitting we used some of the models from Table
\ref{tab:models}, however without the disc component, which could
not be constrained by the PCA data and had little effect above 3
keV. We also fixed the absorption column at $\NH =
1.0\times10^{22}$ cm$^{-2}$.

Model DBTH gave a good total (i.e. summed over all 20
observations) $\chi^2/\nu$ = 2153.7/2440. The spectral parameters
remained roughly constant throughout the outburst. In this model
the soft blackbody component was very weak, its normalization (or
apparent area) consistent with zero in about half of the fits and
its contribution to the bolometric luminosity negligible. On the
other hand its temperature was well constrained, as it was equal
to the temperature of the seed photons in the Comptonized
component and constrained by the low-energy cutoff in the
spectrum.

When we made seed photon temperature a free parameter (model
DBTF), some of the fits become unstable, in particular the
electron temperature was not constrained. We fixed $k\Te$ = 42 keV
(the value taken from the outburst average, see below). After this
modification we obtain total $\chi^2/\nu$ = 2106.4/2240. The
results are presented in Fig.~\ref{fig:pars}. The blackbody and
seed photon temperatures were slightly higher and the spectral
index slightly softer in the beginning of the outburst, but apart
from that the spectral shape remained amazingly constant
throughout the outburst. The ratio of unabsorbed fluxes from
Comptonization and blackbody was $\sim$ 6.

We also fitted the average spectrum of the entire outburst (co-added
observations 2--21). Model DBTF gave a very good fit, $\chi^2/\nu =
101.4/121$. The blackbody temperature was $kT_{\rm bb} = 0.83\pm0.06$, seed
photon temperature $kT_{\rm seed} = 1.46_{-0.18}^{+0.13}$ keV, the spectral
index $\Gamma = 1.93\pm0.03$ and electron temperature $k\Te = 42_{-7}^{+9}$
keV. These parameters were consistent with those obtained from the combined
{\it XMM} and {\it RXTE\/} fits. Compton reflection was not significantly
detected again: the was only an upper limit on $\Omega/2\pi < 0.15$, with fit
improvement of $\Delta\chi^2$ = 2.0 and two degrees of freedom less. This
average outburst spectrum formed a base for analysis of the phase-resolved
spectra in Section~\ref{sec:phase-resolved}.

\subsection{Summary of spectral results}

In the above section we have fitted the X-ray spectrum of \1751 by models
consisting of the emission from accretion disc, blackbody hotspot and
optically thin Comptonization. We have considered two types of models,
corresponding to hotspot/shock geometries sketched in
Fig.~\ref{fig:shockgeom}. In the first geometry the seed photons for
Comptonization originated from the hotspot (DBTH and DBPS,
Fig.~\ref{fig:spec_px}a). In the second geometry the seed photons were
considerably hotter than the blackbody (DBTF and DBPF,
Fig.~\ref{fig:spec_px}b). The second geometry was preferred both from
statistical point of view (it gave better fit) and from the physical
constraints, as the first geometry required the area of the seed photons
source larger than the neutron star surface. In the geometry of the short
shock above the surface, GDB02 ruled out the origin of the seed photons from
the accretion disc.

\section{Pulse profiles}
\label{sec:pulse_profiles}

\begin{figure}
\begin{center}
\leavevmode \epsfxsize=8cm \epsfbox{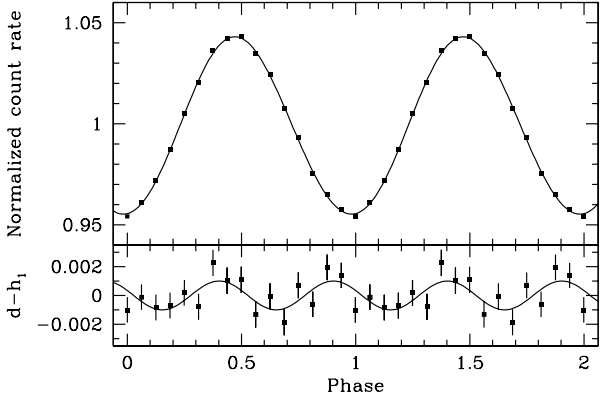}
\end{center}
\caption{Pulse profile of \1751. The upper panel shows the folded
2--60 keV PCA light curve accumulated over the entire outburst.
The errors are of order of symbol size. The curve represents the
model consisting of two harmonics. The lower panel shows the
residuals of data minus first harmonic and the second harmonic
model curve. } \label{fig:profile}
\end{figure}

Fig.~\ref{fig:profile} shows the pulse profile of \1751 accumulated over the
entire outburst. It was not perfectly symmetric, a fit by a single sine
function gave rather poor $\chi^2/\nu$ = 28.3/13. With the second harmonic
added (its phase independent of the first one), the fit was much improved,
$\chi^2/\nu$ = 17.3/11. The rms amplitudes, calculated from these fits, were
3.28$\pm$0.03 and 0.11$\pm$0.03 per cent for the first and second harmonic,
respectively, which correspond  to the peak-to-peak amplitudes
$(F_{\max}-F_{\min})/(F_{\max}+F_{\min})$ of 4.6 and 0.15 per cent.
 The lower panel of Fig.~\ref{fig:profile} shows the residuals of
the single-harmonic fit with the second harmonic function overplotted.

\begin{figure}
\begin{center}
\leavevmode \epsfxsize=8cm \epsfbox{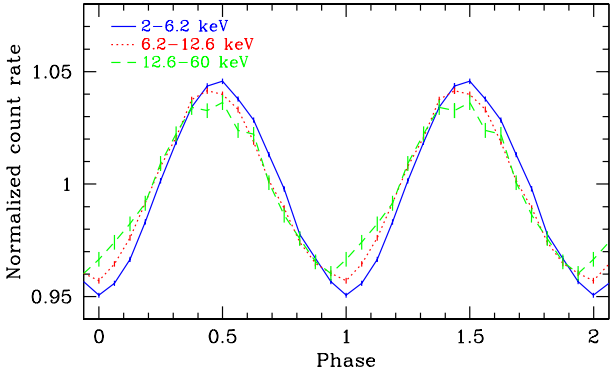}
\end{center}
\caption{Pulse profiles (outburst averaged) in three energy
bands.} \label{fig:prof3}
\end{figure}

\begin{figure}
\begin{center}
\leavevmode \epsfxsize=8cm \epsfbox{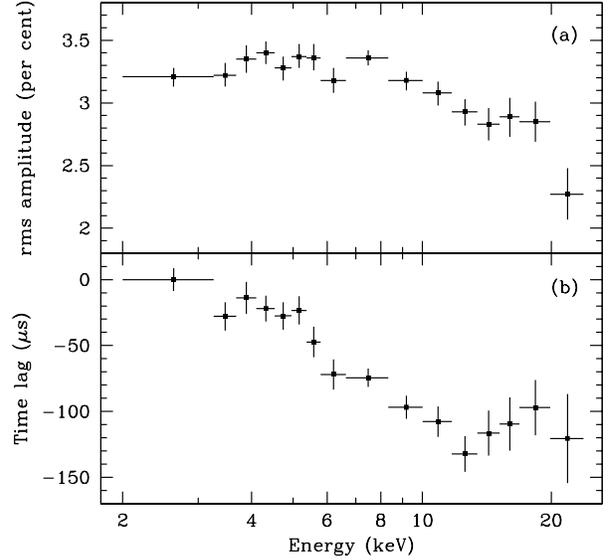}
\end{center}
\caption{(a) The rms amplitude of the pulse profile as a
function of energy. (b) Time lag of the pulse profile versus
energy, with respect to the first energy channel, 2--3.3 keV.}
\label{fig:rms_lag}
\end{figure}

We investigated the energy dependence of the pulse profile.
Fig.~\ref{fig:prof3} shows the outburst-averaged pulse profile in
three energy bands. The pulse amplitude decreased and its profile
shifted towards earlier phase with increasing energy. This effect
can be clearly seen in Fig.~\ref{fig:rms_lag}, where we calculated
the amplitude and phase lag in each of the PCA energy channels by
fitting their pulse profiles by a sine function.
Fig.~\ref{fig:rms_lag}a shows the dependence of the pulse rms
amplitude on energy, which decreased from around 3.3 per cent at
2--10 keV to less than 2.5 per cent at 20 keV. The absolute value
of the time lag with respect to the energy channel 2--3.3 keV
increased with energy up to about 120 $\mu$s at 10 keV, above
which it seemingly saturated (Fig.~\ref{fig:rms_lag}b). The
negative time lag means that the hard photons arrived earlier than
the soft photons. This behaviour was very similar to other
millisecond pulsars, \sax1808 (\citealt*{cmt98}; GDB02),
XTE~J0929--314 \citep{gal02},  and IGR~J00291+5934 \citep{gal05}.

\begin{figure*}
\begin{center}
\leavevmode \epsfxsize=12cm \epsfbox{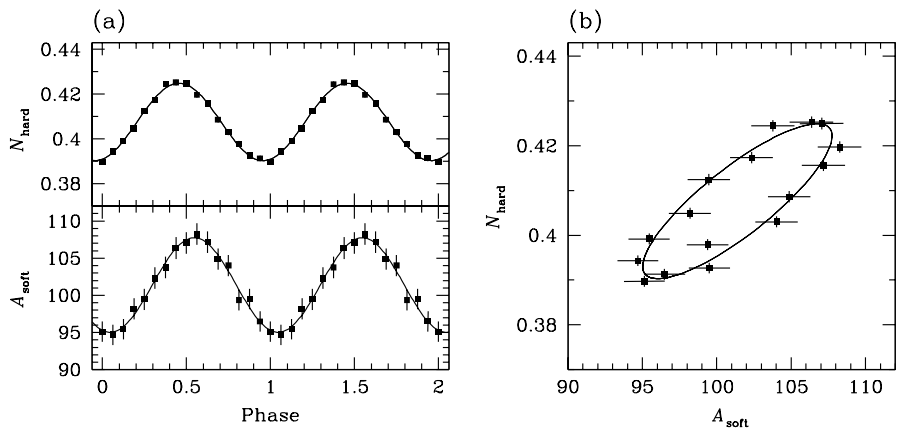}
\end{center}
\caption{Pulsation in the model consisting of the blackbody and
Comptonization of the seed photons independent of this blackbody.
Only normalizations of both components were free during the fits.
This model with total $\chi^2/\nu$ = 501/688 is preferred over
the pulsating index model from Fig.~\ref{fig:fars_bbtg}.}
\label{fig:fars_bbtf}
\end{figure*}

\begin{figure*}
\begin{center}
\leavevmode \epsfxsize=12cm \epsfbox{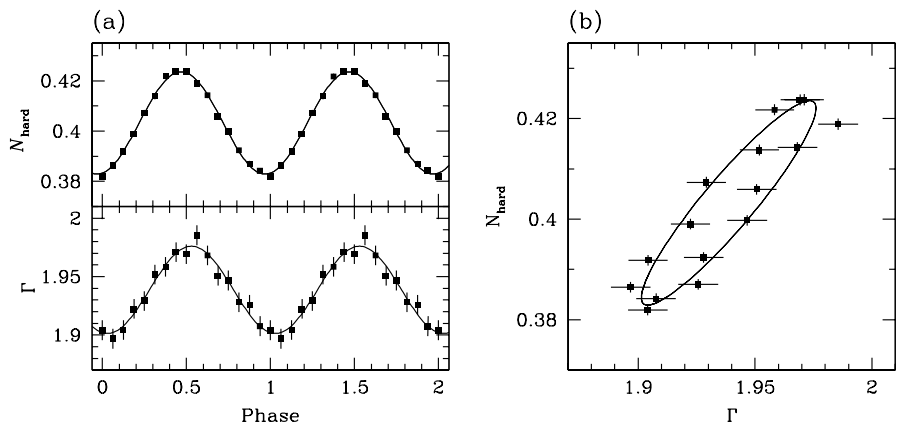}
\end{center}
\caption{Pulsation in the model consisting of the blackbody and
Comptonization of the seed photons independent of this blackbody.
Only normalization of the hard component and its spectral index
were free during the fits. This model gave worse fit, $\chi^2/\nu$
= 592/720, than the two pulsating components model from
Fig.~\ref{fig:fars_bbtf}.}
\label{fig:fars_bbtg}
\end{figure*}

\section{Phase-resolved energy spectra}
\label{sec:phase-resolved}

The photon statistics in the phase-resolved spectra was very
limited, as each of the phase bins contained only a sixteenth part
of the total counts. This made spectral fitting of individual
observations difficult. However, as we have shown in the previous
section, the spectral shape remained almost constant throughout
the outburst, so analysis of the outburst-averaged data was
feasible. Thus, we fitted the average phase-resolved spectrum in
each of the phase bins by the models developed earlier in this
paper.

The best physically motivated models for the broad-band data were
DBTF and DBPF, from which we chose DBTF as it is much faster to
compute. We simplified this model by removing the disc component,
which is not constrained by the PCA-only data. We fixed $\NH =
1.0\times10^{22}$ cm$^{-2}$ and $k\Te$ = 42 keV. This, however,
proved to be problematic. Though the total $\chi^2/\nu$ (i.e.
summed over all 16 phase bins) was very good (407/672), there
was a strong correlation between the spectral index and seed
photon temperature in the limited PCA bandpass. When both
parameters were allowed to be free, the fits tended to give very
low $T_{\rm seed}$, hard $\Gamma$, and much higher $T_{\rm soft}$
than in the broad-band fits. These results were inconsistent not
only with the EPIC-pn data, but also with the PCA/HEXTE phase- and
outburst-averaged spectrum. Therefore, we forced softer spectral
index, and fixed $\Gamma = 1.93$, the value obtained from the
PCA/HEXTE outburst-average fit (Section \ref{sec:xte_fits}). This
gave worse overall $\chi^2/\nu$ = 480/688, but the fitting
parameters were consistent with broad-band fits.

We followed GDB02 in the fitting procedure to eliminate spectral parameters
that did not pulsate significantly, i.e. for each the intrinsic variance [eq.
(1) in GDB02] was zero. In each step of the fitting procedure we fixed one of
the non-pulsating parameters at its mean value from the previous fit, until we
ended up with just two pulsating parameters: normalization of the blackbody
(its apparent area) and normalization of {\sc thcomp}. The total $\chi^2/\nu$
was 501/720. The result is presented in Fig.~\ref{fig:fars_bbtf}. The
normalization of the Comptonized component can be well fitted by the profile
\begin{displaymath}
  N_{\rm hard}(\phi) = 0.41 + 0.017 \sin[2 \pi (\phi - 0.21)] ~ 10^{-9}
  {\rm erg\ s}^{-1} {\rm cm}^{-2},
\end{displaymath}
where $\phi$ is the phase. The blackbody apparent area can be
described by
\begin{displaymath}
  A_{\rm soft}(\phi) = 101 + 6.4 \sin[2 \pi (\phi - 0.30)] ~ {\rm km}^2.
\end{displaymath}
The intrinsic variance of $N_{\rm hard}$ and $A_{\rm soft}$ was
$3.1\pm0.1$ and $4.4\pm0.4$ per cent, respectively. The second
harmonic was not significantly present in any of these profiles,
with upper limits of 8 and 12 per cent of the first harmonic, in
$N_{\rm hard}$ and $A_{\rm soft}$, respectively.

Next, we checked if the spectral variability with phase can be reproduced not
by two independently pulsating components, but by change in the spectral index
of Comptonization. We modified the above spectral fits, starting with the
model where $\Gamma$ was free but the seed photon temperature was fixed
$kT_{\rm seed} = 1.46$ keV (Section \ref{sec:xte_fits}). In the last-but-one
step three parameters were allowed to vary: $A_{\rm soft}$, $N_{\rm hard}$ and
$\Gamma$. Both normalizations pulsated with non-zero intrinsic rms, while the
intrinsic rms of $\Gamma$ was zero. When we fixed $A_{\rm soft}$ and allowed
only $N_{\rm hard}$ and $\Gamma$ to vary, we obtained the overall fit with
$\chi^2/\nu = 592/720$, much worse than in the previous case (see Fig.
\ref{fig:fars_bbtg}). Clearly, the model with two components pulsating
independently was preferred over the pulsating spectral index.

In reality, these  two approaches may not be separated.
In Comptonization models, the angular distribution
of the escaping radiation depends on the scattering order
\citep[see e.g.][and Fig.~\ref{fig:angint}]{st85,vp04}. Thus the spectral
index of Comptonized radiation is a weak function of the
viewing angle, therefore, a slab of Comptonizing plasma on the surface of
the spinning star can produce a spectrum with pulsating spectral index
(as in the second approach). At low energies,
one expects a contribution from the seed, blackbody photons
which, being a strong function of the inclination $\exp(-\tau/\cos i)$,
can give rise to the pulsations of  relative normalizations of
soft and hard spectral component (as in the first approach).
To check  this,  we have fitted the
phase-resolved spectra with {\sc compps} model, which produces angle-dependent
spectrum, where two parameters, inclination and normalization,
were allowed to vary.
The fits occurred to be rather unstable, producing multiple
minima in $\chi^2$ for small and large inclinations. The high-angle
solution yields the mean slab inclination of about $68\degr$ and
amplitude of only about $2\degr$.
Such a small variation of $i$ would be possible if
the shock (and magnetic pole) were almost perfectly aligned with
the rotational pole of the star.

\subsection{Summary of phase-resolved results}

In the section above we have applied two distinctly different approaches to
the phase-resolved spectra. In the first approach there were two components of
fixed spectral shape (blackbody and Comptonization), varying independently
with phase. In the second approach the spectral slope of the hard component
(Comptonization) varied as a function of phase. The crucial common feature of
these two models was the lag in pulsation between softer and harder energies.
This manifested itself as energy-dependent time lags (Fig.~\ref{fig:rms_lag}b)
where the hard photons arrive earlier than the soft photons. The model with
two independently varying components was preferred over the varying index.
In reality, both models can operate together.

\section{Power density spectra}
\label{sec:pds}

\begin{figure}
\begin{center}
\leavevmode \epsfxsize=7cm \epsfbox{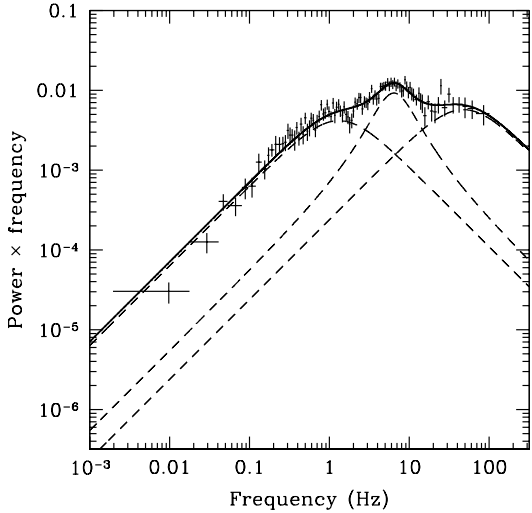}
\end{center}
\caption{Power density spectrum of observation 2 from Table
\ref{tab:obslog} fitted by a multi-Lorentzian model.}
\label{fig:pds}
\end{figure}

Multiple Lorentzian profiles can give a good description of the
power density spectra (PDS) of black hole \citep[e.g.][]{n00}  and neutron star
binaries \citep[e.g.][]{vstra02}. We followed this
approach and fitted the PDS of all observations from Table
\ref{tab:obslog} by a model consisting of up to three Lorentzians.
To avoid the uncertainties of the high-frequency part of the
spectrum we limited the fit to the frequency band of 0.01--100 Hz.
An example of the PDS with the best-fitting model is presented in
Fig.~\ref{fig:pds}.

\begin{figure}
\begin{center}
\leavevmode \epsfxsize=7cm \epsfbox{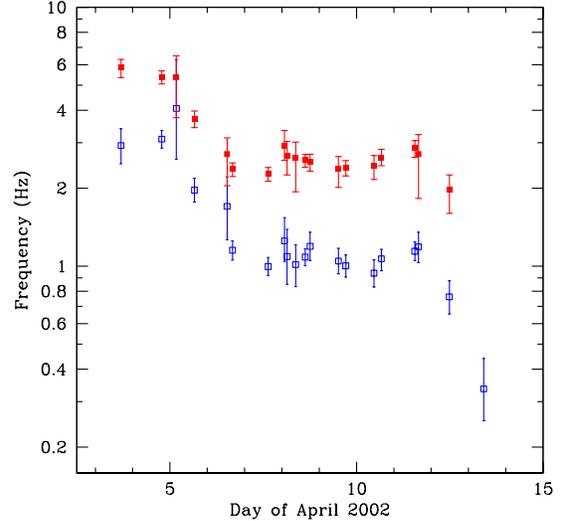}
\end{center}
\caption{Evolution of characteristic frequencies in the PDS
throughout the outburst. The open squares correspond to the width
of the low-frequency, zero-centred Lorentzian, $\nu_{\rm LF}$. The
filled squares describe the QPO frequency, $\nu_{\rm QPO}$.}
\label{fig:freq}
\end{figure}

The low- and high-frequency Lorentzians, describing the broad
noise components were zero frequency-centred. The mid-frequency
Lorentzian function had its centre frequency free. The
characteristic frequency (width) of the low-frequency Lorentzian,
$\nu_{\rm LF}$, can be attributed to the low-frequency break in
the PDS. The centre frequency of the mid-frequency Lorentzian,
$\nu_{\rm QPO}$, corresponds to the quasi-periodic oscillation
(QPO) at 2--6 Hz. Evolution of these two frequencies throughout
the outburst is shown in Fig.~\ref{fig:freq}.

\section{Discussion}
\label{sec:discussion}

\subsection{Comparison with \sax1808}

GDB02 built a physically-motivated model of another millisecond
pulsar, \sax1808, analysing its phase-averaged and phase-resolved
X-ray spectra. According to this model the accretion disc is
disrupted at some distance from the neutron star by its magnetic
field, and collimated towards the poles to create an accretion
column impacting on to the surface. The material in the column is
heated by a shock to temperatures of $\sim$40 keV. The hotspot on
the surface with temperature $\la$ 1 keV provides seed photons for
Comptonization in the hot plasma. Due to a different character of
emission from the optically thick spot and optically thin shock
there is a phase shift between the soft and hard photons. PG03
extended this model to incorporate relativistic effects and
different angular distribution of emitted radiation from the hotspot
and the shock. They showed that the emission patterns
expected from the optically thick blackbody and optically thin
Comptonization can explain the observed phase-resolved spectra
indeed.

The X-ray properties of \1751 are remarkably similar to \sax1808.
The outburst light curve has an alike exponentially decayed
profile \citep{g98} with a break or re-flare \citep*{sck98}
 about 13 days into the outburst. The 3-200
keV spectral shape, entirely typical of an atoll source in the
hard/island state \citep[e.g.][]{bv94,gd02},
is very similar to \sax1808, and likewise almost
constant during the outburst (\citealt*{g98}; GDB02). Both
sources showed coherent pulsations at around 400 Hz, with similar
rms amplitude of a few per cent \citep{wk98a}
and similar energy-dependent soft time lags \citep{cmt98}. The
power density spectra below 100 Hz have similar shape and
frequencies of the QPO and break in the spectra correlated in a
similar way \citep{wk98b}.

The peak outburst luminosity of \1751 was about an order of
magnitude greater than in \sax1808, but the distance to the former
source is uncertain (see discussion in Section
\ref{sec:distance}).

The main difference in spectra between \sax1808 and \1751 was the
weaker blackbody component in the latter source. This can be
easily explained by larger optical depth of the hot plasma, if the
blackbody hotspot is covered by the Comptonizing shock. As GDB02 used
models in different plasma geometry than in this paper, it was
difficult to compare them directly. Therefore, we have fitted the
outburst average PCA/HEXTE spectrum of each source with the same
model consisting of a blackbody and its Comptonization ({\sc
compps}) in slab geometry, inclined at 60$^\circ$ (corresponding
to DBPS model without the disc). We found $k\Te = 60_{-9}^{+7}$
keV and $\tau = 0.88_{-0.04}^{+0.10}$ for \sax1808 and $k\Te =
33_{-2}^{+3}$ keV and $\tau = 1.7\pm0.1$ for \1751. Taking into
account the inclination of the slab, the line-of-sight optical
depth was 1.8 and 3.4, and the amount of unscattered seed photons
reaching the observer was 17 and 3 per cent, respectively.
One can note here that the product $\tau \Te$ is almost the same
in the two sources. This is natural in the two-phase models
(cool neutron star plus a hot shock above), where
the temperature depends on the optical depth, because of the
energy balance constraints
\citep*[see e.g.][]{hm93,stern,ps96,mbp01}.

The pulsation of the soft component in \1751 was significantly
smaller (intrinsic variance of 4.4 per cent) than in \sax1808
(18.1 per cent), though we must stress that different models were
used to estimate these numbers. The pulse profile of the hard
component was less skewed than in \sax1808. The upper limit on the
second harmonic was 8 per cent of the first harmonic, while
\sax1808 required the second harmonic at 18 per cent level.
The relative weakness of the second harmonic in \1751 is expected,
since the ratio of the amplitudes of the second to the
first harmonic is proportional to the  product
$\sin i\sin\theta$ \citep{p04}, where $i$ is the inclination of the
rotational axis and $\theta$ is the co-latitude of the hotspot.
At the same time, the variability amplitude is also proportional to the
same product (see eq.~[\ref{eq:auq}] below).
Thus the smaller is the observed variability amplitude,
the closer the profile is to a sinusoid.

Phase lags between the soft and hard photons were also smaller in
\1751 (see Fig. \ref{fig:rms_lag}b in this paper and fig. 4 in
\citealt{cmt98}). In both sources the time lags are negative, i.e.
hard X-rays arrive before soft X-rays. The energy dependence was
also similar. The absolute value of phase lags increased with
energy up to $\sim$10 keV, above which they saturate at different
levels: $\sim$ 200 $\mu$s (8 per cent of the pulse period) for
\sax1808 and $\sim$ 100 $\mu$s (4 per cent) for \1751. Smaller
values of the phase lags can result from a weaker soft component
-- the black body has a very different radiation pattern from the
Comptonized emission resulting in a different pulse profile
\citep[see][]{vp04}.

There was no Compton reflection significantly detected in any of
the models used in this paper, while GDB02 reported weak
($\Omega/2\pi \sim 0.1$) but statistically significant reflection
in \sax1808. On the other hand, the upper limit on the reflection
amplitude from {\it RXTE} data only (as used by GDB02) was 0.15,
making it consistent with \sax1808.

\subsection{Spectral components}
\label{sec:spectr}

\subsubsection{Origin of the soft components}

The best-fitting model to the broadband data required three
components (Fig.~\ref{fig:spec_px}): two soft components
(`blackbodies') and the hard one (`power law'). We interpreted
them as emission from the accretion disc (low-temperature soft
component), the hotspot on the neutron star surface
(high-temperature soft component) and Comptonization in the shock
above the stellar surface (the hard component).

The simplest argument for this interpretation of soft components
came from their apparent area. We can easily compare them directly
by replacing the disc component in DBTF model with another
single-temperature blackbody, so the model would consist of two
blackbodies and Comptonization. We fitted this model to the
broadband spectrum and obtained a good fit. As a result, the
apparent area of the low-temperature blackbody was $\sim$10 times
larger than the hotter one.

Furthermore, we would not expect any pulsation from the disc, as it rotates
independently of the neutron star. Since the reflection is low, irradiation of
the disc by the spinning spot/shock is negligible. The soft component in the
PCA band (the hotter soft component from Fig.~\ref{fig:spec_px}) showed clear
pulsation with about 4 per cent rms, so it did not originate from the disc. We
can also rephrase this argument in a model-independent way. If the disc
emission had strongly contributed to the PCA spectrum, we would see drop in
the pulse rms at soft X-rays, but it was not the case
(Fig.~\ref{fig:rms_lag}a). Therefore, the disc emission must be weak above 3
keV, and the observed hotter blackbody rather comes from the neutron star.

\subsubsection{Origin of the hard X-ray emission}

Since the hard X-ray emission is pulsed, a fraction of it {\em
must} originate from regions confined by the magnetic field. The
most obvious source of hard X-rays is the place where material
collimated by the magnetic fields impacts on to the surface
\citep{bs76}. However, the pulsed emission can also be
released in the magnetosphere further away from the surface
\citep[e.g.][]{ak90}. In addition, an unknown fraction of the X-rays
can come from the corona above the accretion disc \citep*{grv79}.

The X-ray spectrum of \1751 was fairly constant as a function of
pulse phase and our best-fitting model consisted of two components
of constant spectral shape pulsating independently. This favours
geometries where all of the emission comes from the same source,
as opposed to the sum of separate constant and pulsed components.
The idea of the bulk of emission originating from polar caps is
also supported by the observed broadening of the pulse peak in the
PDS in \sax1808 due to modulation of the aperiodic variability by
the harmonic spin period \citep{men03}. We cannot yet confirm this
result for \1751, but the analogy between the two sources makes a
strong argument for the emission predominantly coming from the
short, heated shock in the polar region.

The constancy of the spectral slope  during
the whole outburst can be used as an argument that the
emission region geometry does not vary much with the accretion rate.
If the  energy dissipation takes place in a hot shock, while the cooling
of the electrons is determined by the reprocessing of the hard
X-ray radiation at the neutron star surface (two-phase model,
\citealt{hm93,stern,ps96,mbp01}),
the spectral slope is determined by the energy balance in the hot phase
and, therefore, by the geometry.

Our spectral models in Section \ref{sec:pn-xte_fits} were
consistent with the slab geometry. For the two models with {\sc
compps} we assumed a slab of hot plasma irradiated from below by a
blackbody. The seed photon area can be determined from the
Comptonization theory because  the amplitude of the Comptonized
component depends linearly on the seed photon flux.
In model DBPS (see Table \ref{tab:px_fits}) the seed photons came from the
blackbody component of the spectrum, with temperature of $\sim$
0.8 keV. This would correspond to the emission region geometry
depicted in Fig.~\ref{fig:shockgeom}(a). However, the area of the
seed photons in this model, i.e. the spot under the slab, was very
large, $A_{\rm seed} \sim$ 800 km$^2$, for a distance of 8.5 kpc
(see discussion of distance in Section \ref{sec:distance}). In the
best-fitting model DBPF the seed photons were independent of the
blackbody component. This could correspond to a spot with a hotter
centre and cooler edges (Fig.~\ref{fig:shockgeom}b). The central
part of the spot with temperature of $\sim$ 2 keV was covered by
the Comptonizing shock and was not visible directly. This
constituted the main source of the seed photons. The cooler
($\sim$ 1 keV) outer part of the spot could be seen directly. The
apparent area of the inner and outer spot were $A\sim$ 20 and
$\sim$ 90 km$^2$, respectively, assuming distance of 8.5 kpc.

\subsection{Geometry}
\label{sec:geometry}

\subsubsection{Inner radius of the accretion disc}
\label{sec:innerdisc}

The spectral models from Section \ref{sec:pn-xte_fits} predicted
the disc to be truncated fairly close to the neutron star, at
$R_{\rm in} \sqrt{\cos i} \ga$ 10 km. Our preferred model DBPF
yielded $R_{\rm in} \approx 14$ km for the inclination of
60$^\circ$. The inner disc radii from Table \ref{tab:px_fits} were
calculated assuming the torque-free inner boundary condition, so
the disc temperature was: \be \label{eq:disctemp} T^4_{\rm eff}
\propto R^{-3} \left(1 - {\sqrt{R_{\rm in} \over R}}\right). \ee
The optically thick disc might be truncated into an inner hot
flow, as it has been suggested for black holes
\citep*[e.g.][]{sle76,emn97}, or
disrupted by the star's magnetic field. In either case it is
possible that the torque is retained through the transition region
\citep[e.g.][]{ak00}. Dropping the inner boundary condition
term from Eq. (\ref{eq:disctemp}) would have increased the derived
$R_{\rm in}$ by factor $\approx 2.7$ \citep{g99},
in which case DBPF would have given $R_{\rm in} \approx 40$ km.

Another important factor is the uncertainty in the distance to
\1751. The above values of $R_{\rm in}$ were calculated for the
distance of $D$ = 8.5 kpc. If the source were at 3 kpc (see
Section \ref{sec:distance}), we would have $R_{\rm in}$ $\approx$
5 and 13 km from DBPF model with torque-free and torqued boundary
condition, respectively.

\begin{figure}
\centerline{\epsfig{file=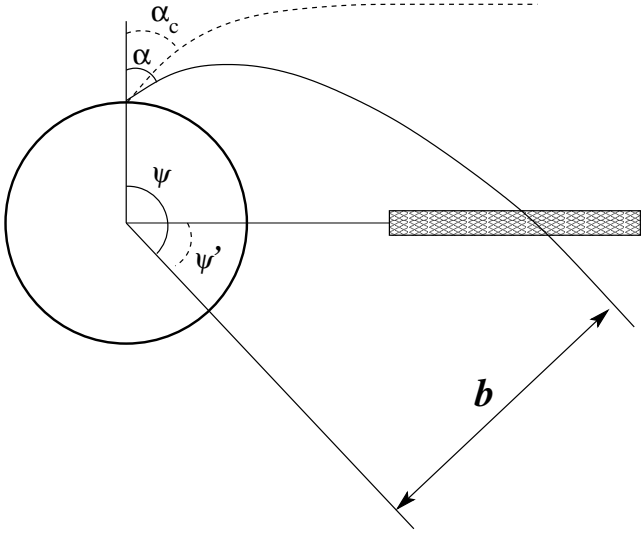,width=8.5cm} }
\caption{Trajectory of a photon in the gravitational field of a
neutron star. Photons emitted at angle $\alpha>\alphac$ relative
to the stellar normal cross the disc plane.
}
\label{fig:trajectory}
\end{figure}

We can estimate the inner disc radius independently from the
observed amplitude of reflection. Let us assume that the emission
originates from a small spot close to the rotational pole of the
neutron star. (These assumptions give a lower limit on reflection
amplitude.) Because of the light bending, a photon   emitted at an
angle $\alpha$ relative to the rotation axis, travels at the
infinity along the trajectory which makes an angle $\psi$ with the
axis (see Fig.~\ref{fig:trajectory}). The relation between
$\alpha$ and $\psi$ is given by an elliptical integral
\citep*{pfc83}. At small $\alpha$, photons propagate in the upper
hemisphere never crossing the disc plane. A photon emitted at a
critical angle $\alphac$ flies parallel to the disc surface. A
simple estimate for this angle can be obtained using
\citet[][hereafter B02]{b02} approximation for light bending:
\be\label{eq:bend} \cos\alpha=\rg/R_*+(1-\rg/R_*)\cos\psi , \ee
where $R_*$ is the stellar radius and $\rg\equiv 2GM/c^2$ is the
Schwarzschild radius for a star of mass $M$. Substituting
$\psi=\pi/2$, we immediately get $\cos \alphac=\rg/R_*$. Photons,
emitted at an angle $\alpha$ satisfying condition
$0<\cos\alpha<\rg/R_*$, will cross the disc plane at a radius
given by an approximate expression for the photon trajectory (B02)
\be \label{eq:rdisk} \frac{R_{\rm disc}(\alpha)}{\rg} \approx
\left[ \frac{ (1-\cos\psi')^2 }{4 (1+\cos\psi')^2 } + \frac{ b^2
}{ \sin^2\psi' } \right]^{1/2} - \frac{
1-\cos\psi'}{2(1+\cos\psi') }, \ee where the impact parameter
$b=R_*\ \sin\alpha/\sqrt{1-\rg/R_*}$, $\psi'=\psi-\pi/2$, and
$\psi$ is related to $\alpha$ via equation (\ref{eq:bend}). The
smallest radius where photon trajectories can cross the disc is
obtained by substituting $\alpha=\pi/2$,
$\sin\psi'=-1/(R_*/\rg-1)$ and $b=R_*/\sqrt{1-\rg/R_*}$\ to
equation (\ref{eq:rdisk}).

We now can compute the reflection amplitude as a function of the
inner disc radius. Let us assume for simplicity that  radiation
from the spot has a constant specific intensity
$I(\alpha)=I_0=const$ (as for the black body emission). We can
compute the luminosity (from the unit area) escaping to the
infinity  as \be \label{eq:Lout} L_{\rm out} = 2\pi
\int_0^{\alphac} I_0 \cos\alpha\ \sin \alpha\ \rmd \alpha= \left[
1- \left( \frac{\rg}{R_*}\right)^2 \right] \frac{I_0}{2} . \ee If
the inner disc radius is larger than $R_{\rm disc}(\pi/2)$, we
reverse relation (\ref{eq:rdisk}) to find the emission angle
$\alpha_{\rm in}$ at which a photon should be emitted to cross the
disc at radius $R_{\rm in}$. The reflection luminosity is then \be
\label{eq:Lrefl} L_{\rm refl} = 2\pi \int_{\alphac}^{\alpha_{\rm
in}} I_0 \cos\alpha\ \sin \alpha\ \rmd \alpha= \left(
\cos^2\alphac - \cos^2\alpha_{\rm in}  \right) \frac{I_0}{2} . \ee
If  $R_{\rm in}< R_{\rm disc}(\pi/2)$, the reflection luminosity
is maximal $L_{\rm refl, max}= (\rg/R_*)^2 I_0/2$.

\begin{figure}
\centerline{\epsfig{file=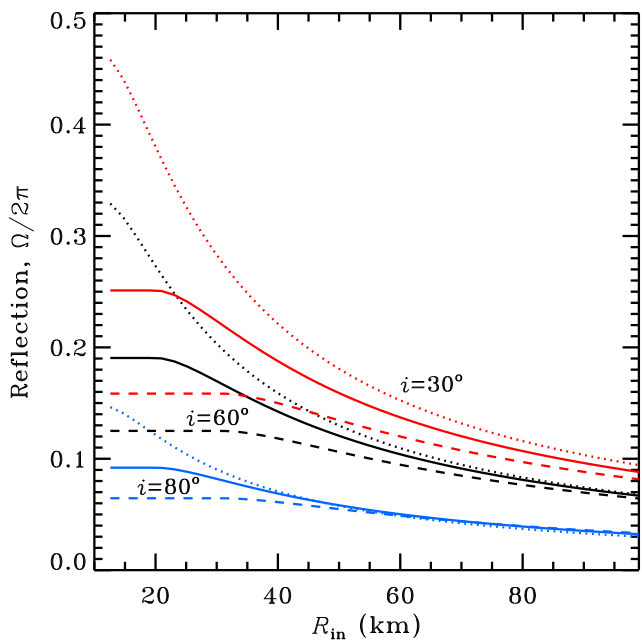,width=8.5cm} }
\caption{Reflection amplitude as a function of the inner disc
radius. The source of photons is a small spot at the rotational
pole of the neutron star. The intrinsic radiation pattern is
assumed to be $I(\mu)=const$ (as for the black body emission).
Dotted curves correspond to $R_*=2\rg$, solid curves to
$R_*=2.5\rg$ and dashed curves to $R_*=3\rg$. The upper curves are
for the inclination $i=30\degr$, middle ones for $i=60\degr$, and
the lower curves are for $i=80\degr$. Neutron star mass of 1.4
$\msun$ is assumed. } \label{fig:refl}
\end{figure}

In order to estimate the reflection amplitude, we need to specify
the angular dependence of the observed escaping and reflected
luminosities. The gravitational redshift affects both luminosities
the same way and thus can be neglected. The escaping luminosity
depends on the inclination angle as (see eq.~[\ref{eq:bend}])
$L_{\rm out}(i) \propto
{\rg/R_*+(1-\rg/R_*)\cos i}$, where normalization factor is found
from the condition $L_{\rm out}=\int_0^1 L_{\rm out}(i) \rmd (\cos
i)$. We can assume that the reflected luminosity is proportional
to the disc area  projected on the sky: $L_{\rm refl}(i)= L_{\rm
refl} 2 \cos i$. Thus  the reflection amplitude is: \be \frac{
\Omega}{2\pi} =\frac{ L_{\rm refl}}{ L_{\rm out} }
\frac{(1+\rg/R_*)\cos i}{\rg/R_*+(1-\rg/R_*)\cos i} \ . \ee It is
plotted in Fig.~\ref{fig:refl} for different stellar compactnesses
and inclination angles. Because the observed upper limit on the
reflection amplitude is $\Omega/2\pi<0.07$ (see
Section~\ref{sec:pn_fits}),  the inner disc cannot be closer than
about 90 km for inclination smaller than $60\degr$. At
$i=80\degr$, constraints become much weaker and depend on stellar
compactness. If stellar radius $R_*<2.5\rg=10.5$ km (neutron star
mass of $1.4\msun$ was assumed), then  $R_{\rm in}>40$ km, but if
$R_*=3\rg$ reflection is smaller (since bending is weaker) and any
inner radius would satisfy these constraints.

\citet{men03} estimated the magnetospheric radius for \sax1808 to
be $\sim$17 km and argued that the accretion disc is disrupted at
this radius. \citet*{vkw04}
analysed QPO frequency correlations in \sax1808 and found that the
upper kilohertz QPO has a frequency by factor $\sim$1.5 lower when
compared to other atoll sources, while low-frequency QPOs were
comparable. They suggested that stronger magnetic field of
\sax1808 moved the inner edge of the disc further away, and
decreased the Keplerian frequency, responsible for the upper kHz
QPO, while low-frequency QPOs were formed further out, and were
not affected. There was no similar frequency shift in \1751
\citep{vkw04}, suggesting weaker magnetic field and smaller
disruption radius than aforementioned 17 km.

Our results about the inner disc radius are not conclusive. The
direct disc emission observed in the broad-band spectrum
corresponded to the disc truncated close to the neutron star. This
result depended on the assumed disc model and a radius of $\sim$
40 km was possible for the disc without the torque-free inner
boundary condition. Similar or larger radius was estimated from
the rather small amount of Compton reflection from the disc. If
the inner disc radius had been larger than the magnetospheric
radius, it would have been truncated into an optically thin hot
inner flow, similarly to the low/hard state of black holes
\citep{sle76,emn97}. We would like to present
an argument in favour of this scenario.

The low-frequency QPO and a break in the power spectrum are likely
to be related to the truncation radius of the disc. The shape of
the low-frequency power spectrum and frequency correlations are
very similar in black holes and neutron stars \citep[e.g.][]{wk99},
so their origin should be independent of the
neutron star surface and magnetic field. The lack of correlation
between the break/QPO frequency and the luminosity (see Figs.
\ref{fig:lightcurve} and \ref{fig:freq}) supports this idea. If
the disc were truncated at the magnetospheric radius indeed, we
would expect its inner radius $R_{\rm in} = R_{\rm magn} \propto
L^{-2/7}$, where $L$ is the luminosity \citep[e.g.][]{h85}. Let
us assume that the low-frequency QPO is connected to the
precession time-scale of a vertical perturbation in the disc
\citep{sv98,pn00}. This yields
$\nu_{\rm QPO} \propto R_{\rm in}^{-3/2}$, so we should expect
$\nu_{\rm QPO} \propto L^{3/7}$. Any other link between the QPO
frequency and the truncation radius would have resulted in the
correlation between $\nu_{\rm QPO}$ and $L$. This has not been
observed. Therefore the optically thick disc was probably
truncated above the magnetospheric radius, by a mechanism
independent of the magnetic field. This could be the same
mechanism truncating the disc of the low/hard state in black hole
binaries, e.g. evaporation \citep[see][]{rc00}. The
low-frequency part of the PDS is formed in this region. Below the
truncation radius the disc is replaced by a hot, optically thin
inner flow, which is eventually disrupted by the magnetic field at
or near the magnetospheric radius.

\subsubsection{Spot size}
\label{sec:spot_size}

Let us estimate the physical size of the emission region using the
model DBPF (hot shocked region surrounded by a cooler black body
edges, see Fig.~\ref{fig:shockgeom}b) which seems physically more
realistic. The inner spot area of $A\sim 20$  km$^2$ corresponds
to the `observed' at the infinity radius of $\rinf=2.5$ km. These
estimations, however, do not take colour hardening and
gravitational redshift and light bending into account.

The observed blackbody (i.e. colour) temperature is different from
the effective temperature by a colour hardening factor, $f_{\rm
col}\equiv T_{\rm col}/T_{\rm eff}$. For the hydrogen and helium
atmospheres of weakly magnetized neutron stars, the exact value of
the colour correction is not very well known, but it has been
estimated to be $f_{\rm col} \sim$ 1.3 \citep*{lvt93,zps96,mjr04}.
This would result in a blueshift of the blackbody spectrum, and
the effective radius of the blackbody would be $R_{\rm spot, eff}
= f_{\rm col}^2 \rinf$. Spectral hardening  results from the fact
that the absorption opacity rapidly decreases with increasing
photon energy and one sees deeper and hotter layers at high
energies. Situation in accreting pulsars can be quite opposite
\citep*[see e.g.][]{dds01}: the hot layers are at the top and
low-energy photons having larger absorption cross-section come
from the hotter outer layers. It is possible that the resulting
colour correction could be negligible, $f_{\rm col}\sim 1$.

Implementing gravitational corrections requires quite complicated
numerical treatment. However, in the case when the (small) spot is
always visible (which is the case for \1751, because we observe
almost sinusoidal oscillations) one can obtain a simple relation
between the observed size at infinity and the physical size at the
neutron star surface $\rspot=\rinf Q^{-1/2}$ (B02; PG03), where
$Q=\rg/R_*+(1-\rg/R_*) \cos i \ \cos\theta$, and $\theta$ is the
 colatitude of the spot centre (magnetic inclination).
The smallest possible radius is obtained for $i=\theta=0$, then
$Q=1$ and $\rspot=\rinf$. This corresponds to the minimum angular
size of the spot of $\rho=12\degr$ ($14\degr$) for $R_*=3\rg$
($2.5\rg$). A reasonable upper limit can be obtained taking
$i=90\degr$ and $\theta=0$ (or  $i=0$ and $\theta=90\degr$), now
$Q=\rg/R_*$ and $\rspot=\rinf \sqrt{R_*/\rg}$. (Note, that for the
isotropically emitting star, the stellar radius is related to the
observed radius as $R_*=\rinf\sqrt{1-\rg/R_*}$, and is {\em
smaller} than that.)

\subsubsection{Constraints from the absence of the secondary maximum}

\begin{figure}
\centerline{\epsfig{file=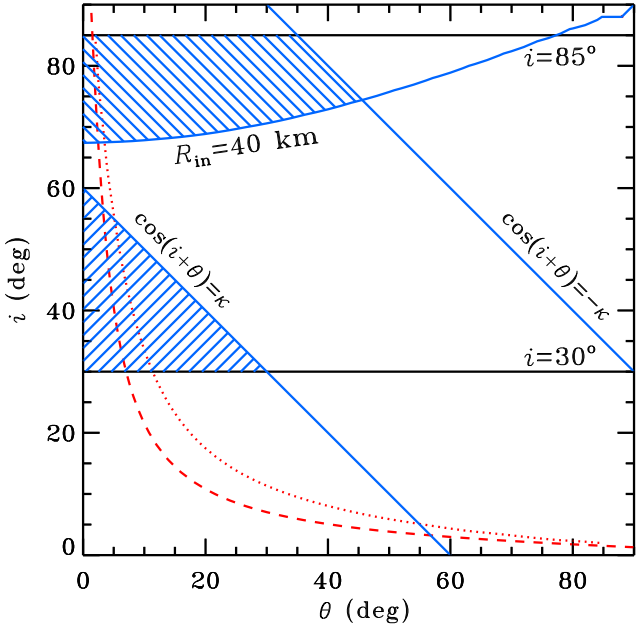,width=8.5cm}} \caption{Constraints
on the inclination and magnetic inclination of the system. The lines
$i=30\degr$ and $i=85\degr$ are the lower and upper limits on the
inclination of the system (M02). A small, $\rho=0$, antipodal spot is
not visible in the region $\cos(i+\theta)>\kappa$ (below the
corresponding line). A finite $\rho$ shifts this line further to the
left by angle $\rho$. The primary spot is eclipsed sometimes when
$\cos(i+\theta)<-\kappa$ (to the right from the corresponding line),
and, therefore, is forbidden. Upper hatched area gives the allowed
regions of parameters $i$ and $\theta$, where the antipodal spot is
blocked by the disc of inner radius $R_{\rm in}=40$ km, and the primary
spot is visible all the time. Contours of constant variability amplitude
$A=0.045$ are shown for a black body spot (dashed curve, computed using
Eq.~\ref{eq:auq}) and for the spot radiation pattern corresponding to
the 5th scattering order in a slab of Thomson optical depth $\taut=1.5$
(dotted curve, numerical simulations for a pulsar of frequency 435 Hz).
Neutron star of mass $M=1.4 \msun$ and radius $R_*=3\rg$ is assumed. }
\label{fig:incldelta}
\end{figure}

The pulse profile of \1751\ is almost sinusoidal. This means
that we see the emission from only one polar cap,
as an antipodal spot would have created a secondary maximum (or a plateau)
in the profile \citep[e.g. B02,][]{vp04}. At the same time, the primary spot
should  never be eclipsed by the star.
 The second  shock can be
obscured by the star or by the optically thick accretion disc.
Let us consider first the possibility of obscuration by the star.


{\it Obscuration by the star}. The condition of obscuration of the
secondary spot (class I in classification of B02) is
$\cos\alpha_{\rm s}<0$. Using analytical formula for light bending
(\ref{eq:bend}) [$\psi=\psi_{\rm s}$ is now the angle between
direction to the observer and normal to the secondary spot,
$\cos\psi_{\rm s}=-(\cos i\ \cos\theta+ \sin i\ \sin\theta\
\cos\phi$), and $\phi$ is the pulsar phase], we get $\cos
(i+\theta)> \kappa$, where $\kappa\equiv\rg/(R_*-\rg)$.  For a
more compact star, $R_*\lesssim 2\rg$ (or about $1.8\rg$ if exact
bending formula is used), such a region does not exist at all.

In Fig.~\ref{fig:incldelta} we show an allowed region
at the $i-\theta$ plane where obscuration is possible.
We restrict the inclination of the system to lie in  the interval
$30\degr< i < 85 \degr$, as argued by M02.
The hatched lower region corresponds to the allowed
space of parameters $i-\theta$.
One sees that the allowed region is rather small.
It becomes even smaller, if we account for the finite spot size.
The obscuration condition is then $\cos (i+\theta+\rho)> \kappa$,
and the boundary of the region shifts further to the left
by angle $\rho$ (i.e.  by $\sim15\degr$),
further diminishing the allowed region.
This region is smaller for more compact stars.
Thus, we find that obscuration of the second spot by the star
is highly improbable, unless the inclination is smaller than $30\degr$
and/or the stellar radius is larger than $3\rg$. In that case, the inner
disc radius has to be at least $\sim200$ km, not to overproduce reflection.

The primary spot is eclipsed sometimes in the
upper right corner of the diagram, $\cos(i+\theta)<-\kappa$ (i.e.
$i+\theta>180\degr-\arccos \kappa$ -- class III in B02 classification),
and, therefore, is forbidden. Again for a spot of finite size,
the limiting line shifts to the left by $\rho$.


{\it Obscuration by the disc}. Let us consider now obscuration of
the secondary spot by the accretion disc. First, we need to
compute radii $R_{\rm disc}$ at which photon orbits (reaching the
observer) cross the disc plane.  One can easily show that in flat
space \be R_{\rm disc}> R_* \sqrt{1+\cos^2\theta/\cos^2i} . \ee In
Schwarzschild metrics, photon orbits cross the disc at somewhat
larger radius due to the light bending. We computed numerically
photon trajectories for every pulsar phase and found the minimum
disc crossing radius. Now we can determine the region in the
$i-\theta$ plane where this radius is larger than the disc inner
radius which, as we argued above, is $R_{\rm in}\sim 40$ km (see
Section~\ref{sec:innerdisc}). We show this region in
Fig.~\ref{fig:incldelta} (upper hatched region, upper right
boundary is determined by the condition of the visibility of the
primary spot). We see that second spot is not visible only for
inclinations above $\sim 67\degr$ (for $R_*=3\rg$). Larger $R_{\rm
in}$ will further reduce the allowed region.

If we account for the finite size of the spot, the allowed region
shifts to the left by the angular size of the spot. However, the
lower limit on the inclinations is not affected much, because now
it is a rather flat function of $\theta$. We conclude that the
second spot can be blocked by the accretion disc. We obtain
constraints  $i\gtrsim67\degr$ ($i\gtrsim70\degr$) and
$\theta<30\degr$ ($\theta<35\degr$) for $R_*=3\rg$ ($2.5\rg$) for
the spot size of $\rho\sim 15\degr$.

\subsubsection{Constraints from the variability amplitude}
\label{sec:constampl}

\begin{figure}
\centerline{\epsfig{file=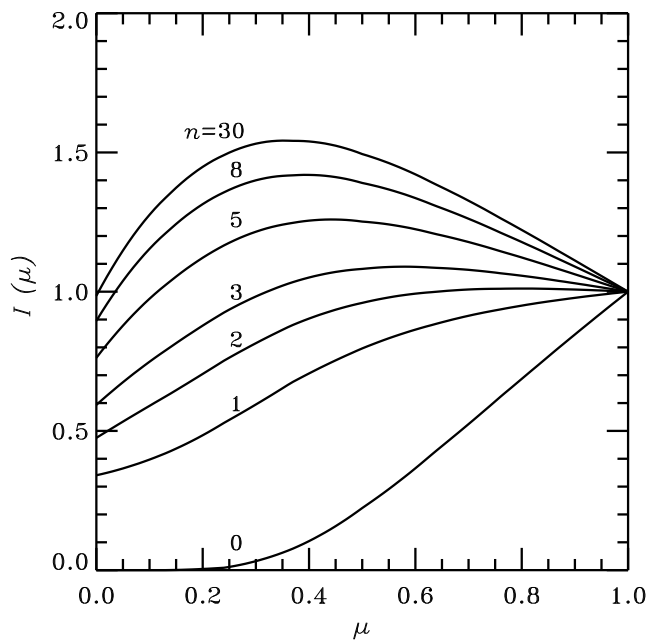,width=8.5cm} }
\caption{Intensity (normalized to unity at $\mu=1$) of the radiation
escaping from a slab of Thomson optical depth  $\taut=1.5$
for different scattering orders $n$ as a function of $\mu$
(here $\mu$ is the cosine of the zenith angle).
The seed photons with intensity $I(\mu)=1$ are injected
from the bottom of the slab.
Calculation are done in Thomson scattering approximation
\citep[see][]{st85,vp04}.
}
\label{fig:angint}
\end{figure}

Important constraints on the physical parameters of the system are
given by the variability amplitude. The observed peak-to-peak
amplitude $A\approx 0.045$ gives the minimum intrinsic variability. Let us
first assume that all the emission comes the spot. For a slowly
rotating star, the  variability amplitude (for a
black body spot) can be estimated from (B02; PG03):
\be\label{eq:auq} A=U/Q, \ee where $U=(1-\rg/R_*)\sin i\
\sin\theta$. The dashed curve in Fig.~\ref{fig:incldelta} shows
the relation between $i$ and $\theta$ satisfying the observed $A$.

In reality, the star rotates rapidly and the radiation pattern from
of the Comptonized emission is far from the black body.
Rapid rotation causes a slight shift of the maximum emission from the
phase where projected area has the maximum towards the phase where
Doppler factor has a maximum (i.e. quarter of the period earlier,
see PG03).
Additionally, light travel time delays skew the profile.

Change in the emission pattern causes even larger changes to the
oscillation profile. The angular dependence of the radiation
escaping from a slab of Thomson optical depth $\taut=1.5$ (as
suggested by spectral fitting, see Section~\ref{sec:pn-xte_fits}
and Table~\ref{tab:px_fits}) is shown in Fig.~\ref{fig:angint}
\citep[see also][]{vp04}. It depends on the scattering order $n$
which is  in turn related to the photon energy as $E/3kT_{\rm
seed}\approx (1+4k\Te/m_{\rm e} c^2)^n$. Thus we expect 3--7 scatterings
to contribute most of the flux in the PCA range. As an example we
compute the light curves expected for  $n=5$ using fully
relativistic method described in PG03 and \citet{vp04}. Contour
plots at the $i-\theta$ plane for $A=0.045$ are shown in
Fig.~\ref{fig:incldelta} by dotted curves. We see that the
Comptonized emission produces smaller variability compared to the
black body, and for the same $i$, larger $\theta$ is needed to get
the same amplitude.

If we take the upper hatched region as the allowed region of parameters,
we immediately get that the magnetic pole is misaligned from the rotational
pole by a very small angle,  $\theta=3-4\degr$.
This angle becomes larger if it is only a fraction of the emission
that comes from the spot.

\subsection{Distance}
\label{sec:distance}

We used the distance of 8.5 kpc throughout this paper, assuming
that the source is close to the Galactic centre.
M02 placed \1751 at a distance of at least 7 kpc, using
indirect arguments. There are, however, clues that this distance
might be actually much smaller.

Firstly, the source is very bright, $L_{\rm peak} = 0.13
\Ledd$, when placed at 8.5 kpc. It is about an order of
magnitude brighter than \sax1808, while all other properties
are very similar. At higher accretion rates, typically more
that a few per cent of the Eddington limit, atoll sources
switch to a soft (banana) spectral state
\citep[e.g.][]{hk89}. This is not the case here. But the
transition luminosity can be quite high, in particular
during the onset of the outburst, where hysteresis effect
has been observed and the source can remain in the hard
state despite high accretion rate \citep{mc03}. For example,
Aql X-1 and 4U 1705--44 have been observed in the hard
island state at luminosities $\sim$ 20 per cent of $\Ledd$
\citep{bo02,dg03}.  Therefore, having \1751 in the hard
state at 13 per cent of the Eddington luminosity cannot be
ruled out.

\citet*{czb01} argued that the crucial difference between the
millisecond pulsars and non-pulsating atolls was in the average
accretion rate (related to the size of the disc, which is smaller
in the millisecond pulsars). In atolls, the accretion rate is high
enough to bury the magnetic field under the surface. The diffusion
time-scale of rebuilding the field is $\sim$1000 years, much
larger than a typical period between the outbursts. In millisecond
pulsars the accretion rate, $\dot{m}$, is never high enough to
bury the field. \citet{czb01} estimated that magnetic screening is
ineffective for $\dot{m} < 0.02 \dot{m}_{\rm Edd}$. Then, 0.13
$\Ledd$ would have been too much for the magnetic field to
survive in \1751 during the outburst. It would  need $D \la 3$ kpc
for $\dot{m}$ to be low enough for the field not to be buried.
This issue would require further study to resolve.

If \1751 were located at $D$ = 3 kpc indeed, it would decrease the
blackbody apparent areas quoted in Table \ref{tab:px_fits} by
factor $(8.5/3)^2 \approx 8$ and disc inner radius by factor
$8.5/3 \approx 3$. The model DBPS (and DBTH), where the seed
photons were from the blackbody component, would give the
blackbody area of $\sim$100 km, and could not be rejected on the
spot size basis. The inner spot area in DBPF model would be
$\sim$3 km$^2$ and its linear size derived in Section
\ref{sec:spot_size} would be less by factor $\sim$3.

\section{Summary}

We have analysed the 2002 outburst of the second accretion-powered
millisecond pulsar \1751. The broad-band 0.7--200 keV spectrum of
\1751 obtained by {\it XMM-Newton} and {\it RXTE} is complex. The
best-fitting models consist of two soft components and thermal
Comptonization. Most likely the cooler soft component comes from
the accretion disc, the hotter one from the surface of the neutron
star, while Comptonization takes place in a shocked region in the
accretion column. We estimate the electron temperature of the
plasma 30--40 keV and Thomson optical depth $\tau\sim 1.5$ (for a
slab geometry). In our best spectral models, the temperature of
the seed soft photons for Comptonization is higher than the
temperature of the visible soft blackbody component which is
consistent with  a picture where   photons from a heated spot
under the shock are scattered away by the shocked plasma. In this
model, the `observed' (at the infinity) area of the blackbody
emission is about 100 km$^2$ corresponding to a spot of radius
$\sim$ 5--6 km. The area corresponding to a shock was estimated to
be 20 km$^2$, which translates to a  radius of 2.5-4.5 km
depending on the assumed compactness of the neutron star and the
viewing angle.

The analysis of the {\it RXTE} spectra obtained during the
outburst shows that temperature of the soft blackbody photons as
well as the temperature of the seed photons decreased
monotonically. The spectral slope of the hard Comptonized
component did not vary much. We argued that this favours a
constant geometry (e.g. slab) emission region and two-phase
models, where energy is dissipated in the hot phase and the seed
photons for Comptonization are the result of reprocessing of the
hard X-rays.

The pulse profile cannot be fitted by a simple sinusoid, a second harmonic is
required. The mean peak-to-peak amplitude of the first and  second harmonic is
about 4.5 and 0.15 per cent, respectively. There is a clear energy dependence
of the profile, the amplitude gradually decreases and the peak at higher
energies appears at earlier phases (soft lags). The time lags reach $\sim 100$
$\mu{\rm s}$ and seem to saturate at about 10 keV. This behaviour is almost
identical to \sax1808 where lags are factor of two larger. The observed energy
dependence of the pulse profiles can be explained by a model where blackbody
and Comptonization components vary sinusoidally with a small phase shift. Such
a behaviour can result from the Doppler boosting and different angular
distribution of the emission from these components (PG03).

The non-detection of the reflected component in the time-averaged
spectra and the absence of the emission from the second antipodal
spot (we see almost sinusoidal variations) put constraints on the
geometry of the system. We argued that the inner radius of the
optically thick accreting disc is about 40 km. In that case, the
secondary can be blocked by the accretion disc if the inclination
of the system is larger than $\sim 70\degr$. Blockage of the
secondary by the star itself is highly unlikely for the neutron
star radii $R_*<3\rg$. The observed variability amplitude
constrain the magnetic pole to lie within 3--4$\degr$ of the
rotational pole, if most of the observed X-ray emission comes from
a hotspot and  a shock.

\section*{Acknowledgements}

This work was supported by the Academy of Finland grants 100488 and
201079, the Jenny and Antti Wihuri Foundation, and the NORDITA Nordic
project on High Energy Astrophysics. We thank the referee for helpful
comments.


\label{lastpage}

\end{document}